\documentclass{aa}  

\usepackage{graphicx}
\usepackage{txfonts}
\usepackage{natbib}
\usepackage{subfigure}
\bibpunct{(}{)}{;}{a}{}{,} 
\usepackage{footnote}
\usepackage{threeparttable}
\usepackage[utf8]{inputenc}
\begin{document} 

  \title{X-ray emission from thin plasmas}
  \subtitle{Collisional ionization for atoms and ions of H to Zn}
  
   \author{I. Urdampilleta
          \inst{1,2}
          \and
          J. S. Kaastra\inst{1,2}
          \and
          M. Mehdipour\inst{1}}

   \institute{SRON Netherlands Institute for Space Research \\
              Sorbonnelaan 2, 3584 CA Utrecht, The Nether\-lands\\
              \email{i.urdampilleta@sron.nl}
         \and
             Leiden Observatory, Leiden University\\
             PO Box 9513, 2300 RA Leiden, The Netherlands}
   \date{Accepted February 16, 2017}

 \abstract{Every observation of astrophysical objects involving a spectrum requires atomic data for the interpretation of line fluxes, line ratios, and ionization state of the emitting plasma. One of the processes that determines it is collisional ionization. In this study, an update of the direct ionization (DI)  and excitation-autoionization (EA) processes is discussed for the H to Zn-like isoelectronic sequences. In recent years, new laboratory measurements and theoretical calculations of ionization cross-sections have become available. We provide an extension and update of previous published reviews in the literature. We include the most recent experimental measurements and fit the cross-sections of all individual shells of all ions from H to Zn. These data are described using an extension of Younger's and Mewe's formula, suitable for integration over a Maxwellian velocity distribution to derive the subshell ionization rate coefficients. These ionization rate coefficients are incorporated in the high-resolution plasma code and spectral fitting tool SPEX V3.0.}
   
   
 


\maketitle  
%

\section{Introduction}

 In calculations of thermal X-ray radiation from hot optically thin plasmas, it is important to have accurate estimates of the ion fractions of the plasma, since the predicted line fluxes sometimes depend sensitively on the ion concentrations. The ion concentrations are determined from the total ionization and recombination rates. In this paper, we focus on collisional ionization rates. Radiative recombination rates \citep{Mao2016} and charge exchange rates \citep{Gu2016} are treated in separate papers. These rates are essential for the analysis and interpretation of high-resolution astrophysical X-ray spectra, in particular for the future era of X-ray astronomy with Athena.   

 An often used compilation of ionization and recombination rates is given by \cite{Arnaud1985}, AR hereafter. AR treat the rates for 15 of the most abundant chemical elements.  Since that time, however, many new laboratory measurements and theoretical calculations of the relevant ionization processes have become available.  A good example is given by \cite{Arnaud1992}, who re-investigated the ionization balance of Fe using new data.  Their newly derived equilibrium concentrations deviate sometimes even by a factor of 2--3 from AR. The most recent review has been performed by \cite{Dere2007}, D07 hereafter. D07 presents total ionization rates for all elements up to the Zn isoelectronic sequence that were derived mainly from laboratory measurements or Flexible Atomic Data \citep[FAC,][]{Gu2002} calculations.

 Motivated by the findings of AR and D07, we  started an update of the ionization rates, extending it to all shells of 30 elements from H to Zn.  Since we want to use the rates not only for equilibrium plasmas but also for non-equilibrium situations, it is important to know the contributions from different atomic subshells separately.  Under non-equilibrium conditions inner shell ionization may play an important role, both in the determination of the ionization balance and in producing fluorescent lines.

 In the following Sect. 2, we give an overview of the fitting procedure used in this work.  In Sect. 3, we review the ionization cross-sections obtained from experimental measurements or theoretical calculations along isoelectronic sequences. Details of the ion rate coefficients analytical approach are given in Sect. 4, Appendix B and C. In Sect. 5 we compare and discuss the results of this work. The  references used for the cross-sections are included in Appendix A.

\section{Fitting procedure}

Collisional ionization is mainly dominated by two mechanisms:  direct ionization (DI), where the impact of a free electron on an atom liberates a bound electron; and excitation-autoionization (EA), when a free electron excites an atom into an autoionizing state during a collision.

\subsection{Direct Ionization cross-sections}

An important notion in treating DI is the scaling law along the isoelectronic sequence, as first obtained by \cite{Thomson1912}:
\begin{equation}
\displaystyle uI^2Q = f(u),
\end{equation}

where $u=E_{\rm e}/I$ with $E_{\rm e}$ (keV) the incoming electron energy and
$I$ (keV) the ionization potential of the atomic subshell; $Q$ ($10^{-24}$~m$^2$) is the
ionization cross-section. The function $f(u)$ does not -- in lowest order -- depend upon the nuclear charge $Z$ of the ion, and is a unique function
for each subshell of all elements in each isoelectronic sequence.

Direct ionization cross-sections are most readily fitted using the
following formula, which is an extension of the  parametric formula originally proposed by \cite{Younger1981a}:
\begin{equation}
\displaystyle uI^2Q_{DI} = A\left(1-{1\over u}\right) + B \left(1 - {1\over u}\right)^2 + C R \ln
u + D {\ln u\over \sqrt{u}} +  E{\ln u\over u}.
\label{eqn:younger_ext}
\end{equation}
The parameters $A$, $B$, $D,$ and $E$ are in units of
$10^{-24}$~m$^2$keV$^2$ and can be adjusted to fit the
observed or calculated DI cross-sections, see Section 2.3 for more details. $R$ is a relativistic
correction discussed below. $C$ is the Bethe constant and corresponds to the high energy limit of the cross-section.

The parameter $C$ is given by \cite{Younger1981c}:
\begin{equation}
C = {IE_{\rm H}\over \pi \alpha} \int {\sigma(E)\over E}{\rm d}E,
\label{eqn:bethe}
\end{equation}
where $\sigma(E)$ is the photo-ionization cross-section of the current subshell,
$E_{\rm H}$ the ionization energy of Hydrogen and $\alpha$ the fine structure
constant.  The Bethe constants used in this paper are derived from the fits to
the Hartree-Dirac-Slater photoionization cross-sections, as presented by \cite{Verner1995}.

As mentioned above,  Eq. \ref{eqn:younger_ext} is an extension of Younger's formula, where we have added the term with  $D {\ln u/ \sqrt{u}}$. The main reason to introduce this term is that in some cases the fitted value for $C$, as determined from a fit over a relatively low energy range, differs considerably from the
theoretical limit for $u\rightarrow \infty$ as determined from
(\ref{eqn:bethe}).  For example, AR give $C$ = 12.0$\times
10^{-24}$~m$^2$keV$^2$ for their fit to the 2p cross-section of C~I, while the
Bethe coefficient derived from (\ref{eqn:bethe}) is 6.0$\times
10^{-24}$~m$^2$keV$^2$.  However, if we fix $C$ to the Bethe value in the fit, the
resulting fit sometimes shows systematic deviations with a magnitude of 10\% of
the maximum cross-section. This is because the three remaining parameters $A$, $B,$ and $E$ are insufficient to model all details at lower energies.  Therefore, we need an extra fit component  which, for
small $u$ is close to $\ln u,$ but vanishes for large $u$,  to accommodate
for the discrepancy in $C$.

The relativistic correction $R$ in (\ref{eqn:younger_ext}) becomes important for large nuclear charge $Z$ (or equivalently large ionization potential $I$) 
and large incoming electron energy $E$ \citep{Zhang1990,Moores1990,Kao1992}. This expression is only valid for the midly relativistic ($\epsilon$~$\lesssim$ 1, where $\epsilon\equiv E/m_{\rm e}c^2$) regime. Our approximations and cross-sections do not apply to the fully relativistic regime  ($\epsilon$~$\gtrsim$ 1). The presence of this correction is clearly visible for the hydrogen and helium sequences, as shown in  Fig.~\ref{fig:Fig2}. Using a classical approach the relativistic correction can be written here as given
by, for example,  \cite{Quarles1976} and \cite{Tinschert1989}:
\begin{equation}
\label{eqn:gryzinski}
R = \left( {\tau + 2 \over \epsilon + 2} \right)
    \left( {\epsilon + 1 \over \tau + 1} \right)^2
    \left[ {
     (\tau+\epsilon) (\epsilon+2) (\tau+1)^2 
     \over
     \epsilon (\epsilon+2) (\tau+1)^2 + \tau (\tau+2)
     } \right]^{3/2},
\end{equation}
where $\tau\equiv I/m_{\rm e}c^2$,
with $m_{\rm e}$ the rest mass of the electron and $c$ the speed of light.
    The above correction factor $R$, when applied to  the simple Lotz-approximation \citep{Lotz1967},
is consistent with the available observational data for a wide range of
nuclear charge values ($Z$=1--83) and 5 magnitudes of energy, within
a range of about 15\% \citep{Quarles1976}.

For the present range of ions up to Zn ($Z$=30), the ionization potential
is small  compared to $m_{\rm e}c^2$ and, hence, $\tau$ is small.
On the other hand, we are interested in the cross-section up to high energies
($\sim$100\,keV) that applies to the hottest thin astrophysical plasmas, and
therefore $\epsilon$ is not always negligible. By making a Taylor's
expansion in $\epsilon$ of (\ref{eqn:gryzinski}) for small $\tau$ we
obtain
\begin{equation}
\label{eqn:gryzinski_approx}
R \approx 1 + 1.5\epsilon + 0.25\epsilon^2.
\end{equation}
We will use this approximation (\ref{eqn:gryzinski_approx}) in our formula
for the DI cross-section (\ref{eqn:younger_ext}).

Analysing the asymptotic behaviour of equation (\ref{eqn:younger_ext})
\begin{equation}
\displaystyle\lim_{u\to1}{uI^2Q_{DI}} = (A + C + D + E) (u - 1), 
\label{eqn:ui2q_low}
\end{equation}
\begin{equation}
\displaystyle\lim_{u\to\infty}{uI^2Q_{DI}} = C\ln u.
\label{eqn:ui2q_high}
\end{equation}

Therefore, it is evident from (\ref{eqn:ui2q_low}) that the fit
parameters $A$ to $E$ must satisfy the constraint $A+C+D+E>0$.  Further, the Bethe constant $C$ gives the asymptotic behaviour at high
energies. 

\begin{figure}
\includegraphics[width=0.52\textwidth]{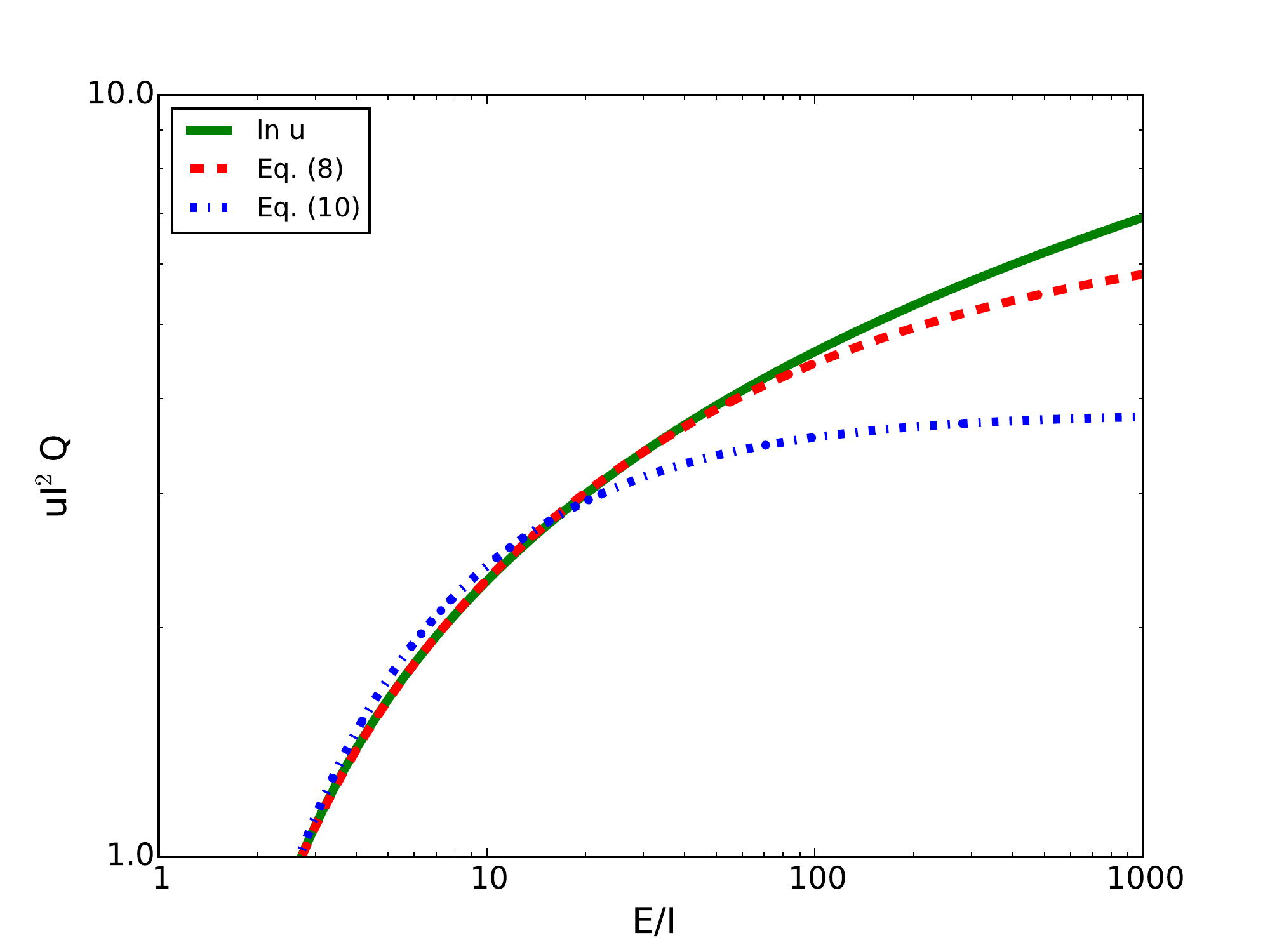}
\caption{Approximation to the Bethe cross-section.}
\label{fig:Fig1}
\end{figure}

It appears that when $u$ is not too large, $\ln u$ can be decomposed as (see Fig. \ref{fig:Fig1})
\begin{eqnarray}
\ln u &\approx 6.5597\left(1-{1\over u}\right) + 0.4407 \left(1 - {1\over u}\right)^2 \nonumber\\
 &-5.3622 {\ln u\over \sqrt{u}} -0.1998 {\ln u\over u}.
\label{eqn:logapprox}
\end{eqnarray}
Equation \ref{eqn:logapprox} has a relative accuracy that is better than 1, 3, and 16\% for $u$ smaller than 50, 100, and 1000, respectively; and the corresponding
cross-section contribution $\ln u/ u$  deviates never more from the true cross-section than 0.5\% of the corresponding maximum cross-section (which occurs at $u = e$).

In all cases, where we do not fit the cross-section, based on Equation \ref{eqn:logapprox}, we use the following expression for the calculation
 of Younger's formula parameter (with 
$A({\rm ref})$, $B({\rm ref})$, $C({\rm ref})$, $D({\rm ref})$ and $E({\rm ref})$), as given by the parameters of the isoelectronic sequence that we use as reference. For example, the Li-sequence is used as reference for the 1s cross-sections of the Be to Zn-sequences as detailed in Section 3.4.1. 
\begin{eqnarray}
A&=&A({\rm ref}) + 6.5597 [C({\rm ref}) - C({\rm Bethe})], \nonumber \\B&=&B({\rm ref}) + 0.4407 [C({\rm ref}) - C({\rm Bethe})], \nonumber \\C&=&C(\rm Bethe), \nonumber \\
D&=&D({\rm ref}) - 5.3622 [C({\rm ref}) - C({\rm Bethe})], \nonumber \\E&=&E({\rm ref}) - 0.1998 [C({\rm ref}) - C({\rm Bethe})].
\label{eqn:interpar}
\end{eqnarray}

This assures that, for most of the lower energies, the scaled cross-section is identical to the reference cross-section, while at high energies it has the correct asymptotic behaviour.

In some cases, we can get acceptable fits with $D=0$. In these cases, we
obtain a somewhat less accurate approximation for the logarithm:
\begin{eqnarray}
\ln u &\approx 6.5867\left(1-{1\over u}\right) - 2.7655 \left(1 - {1\over u}\right)^2 \nonumber\\
 &-5.5528 {\ln u\over u}.
\label{eqn:logapprox2}
\end{eqnarray}

Equation \ref{eqn:logapprox2} has a relative accuracy better than 14, 23, and
45\% for $u$ smaller than 50, 100, and 1000, respectively; and the corresponding
cross-section never deviates  more from the true cross-section than 5.6\% of the
maximum cross-section (which occurs at $u = e$).

In the case of $D=0$, the equivalent of Equation \ref{eqn:interpar} becomes
\begin{eqnarray}
A&=&A({\rm ref}) + 6.5867 [C({\rm ref}) - C({\rm Bethe})], \nonumber \\B&=&B({\rm ref}) - 2.7655 [C({\rm ref}) - C({\rm Bethe})], \nonumber \\C&=&C(\rm Bethe), \nonumber \\
D&=&0, \nonumber \\
E&=&E({\rm ref}) - 5.5528 [C({\rm ref}) - C({\rm Bethe})].
\label{eqn:interpar2}
\end{eqnarray}

\subsection{Excitation-autoionization cross-sections}

The excitation-autoionization (EA) process occurs when a free electron
excites an atom or ion during a collision. In some cases, especially
for the Li and Na isoelectronic sequence, the excited states are often
unstable owing to Auger transitions, leading to simultaneous ejection of one
electron and decay to a lower energy level of another electron. Many
different excited energy levels can contribute to the EA process. In general,  this
leads to a complicated total EA cross-section, showing many
discontinuous jumps at the different excitation threshold energies.
Since in most astrophysical applications we are not interested in the
details of the EA cross-section, but only in its value averaged over a
broad electron distribution, it is reasonable to approximate the true
EA cross-section by a simplified fitting formula.

The EA cross-section is most readily fitted using Mewe's formula, originally proposed to fit excitation cross-sections by \cite{Mewe1972}:
\begin{equation}
u{I_{EA}}^2Q_{EA} = A_{EA}+ B_{EA}/u  + C_{EA} / u^2 + 2D_{EA} / u^3 + E_{EA} \ln u, 
\label{eqn:eamewe}
\end{equation}
where $u\equiv E_{\rm e}/I_{EA}$ with $E_{\rm e}$ the incoming electron
energy; $Q_{EA}$ is the EA cross-section. The parameters $A_{EA}$ to $E_{EA}$ and $I_{EA}$ can be adjusted to fit the observed or calculated EA cross-sections.  We note that \cite{Arnaud1992} first proposed to use
this formula for EA cross-sections, although they used a slightly
different definition of the parameters.

For the Li, Be, and B isoelectronic sequences, we used the calculations of \cite{Sampson1981}. All the necessary formulae can be found in
their paper. The scaled collision strengths needed were obtained from \cite{Golden1978}, Table 5. For these sequences, we used the sum
of two terms with Eq. \ref{eqn:eamewe}, the first  term corresponding to excitation
1s-2$\ell,$ and the second term corresponding to all excitations
1s-$n\ell$ with $n>2$. The total fitted EA cross-section deviates no
more than 5\% of the maximum EA contribution, using the exact
expressions of Sampson \& Golden. Since, for these sequences, the EA
contribution is typically less than 10\% of the total cross-section,
our fit accuracy is sufficient given the systematic uncertainties in
measurements and theory.

For the Na to Ar isoelectronic sequences, AR recommends to extend the calculations for the Na-sequence of \cite{Sampson1982} to the Mg-Ar sequences. In doing so, they recommend to put all the branching ratios to unity. We follow the AR recommendations and use the method described in \cite{Sampson1982}, extended to the Mg-Ar sequences, to calculate the EA cross-section. We consider the branching ratio unity for these calculations. We include excitation from the 2s and 2p subshells to the $n$s, $n$p, and $n$d subshells with $n$ ranging from 3--5.  We then fit these cross-sections to Eq. \ref{eqn:eamewe}, splitting it into two components: transitions towards $n=3$ and $n=4,5$. The advantage of this approach is that we can estimate the EA contribution for all relevant energies. Other theoretical EA calculations are often only presented for a limited energy range.

For the K to Cr isoelectronic sequences we obtain the EA parameters by fitting to Eq. \ref{eqn:eamewe} the FAC EA cross-sections of D07, which are the same used by CHIANTI.

\subsection{Fitting experimental and theoretical data}

The main purpose of our fitting procedure is to obtain the parameters $A$, $B$, $D,$ and $E$ of Eq. \ref{eqn:younger_ext} for all inner and outer shells that contribute to DI process, together with the EA parameters, which are calculated as explained above.

For the H and He-sequences, the DI parameters are obtained directly by fitting the cross-sections from experimental measurements and theoretical calculations listed in Appendix A.
The rest of sequences include the DI contribution of the outer and one or several inner shells. In this case, we cannot perform a direct fit to the data because most papers in the literature only present the total cross-sections, which are not split into subshells, while our purpose is to obtain the individual outer and inner shell cross-sections separately.

For this reason, we calculate first the EA and inner shell DI parameters and cross-sections. The particular method used for each isoelectronic sequence is explained in Sect. 3. Afterwards, to obtain the outer shell cross-section (for example for the Li-like sequence, 2s), we subtract the inner shell (in the Li-like case, 1s) and EA contributions from total cross-section. We then fit this outer shell contribution using Eq. \ref{eqn:younger_ext} and obtain the parameters $A$, $B$, $D,$ and $E$.

The remaining cross-sections, for which no data are present are obtained, using Eq. \ref{eqn:younger_ext} with interpolated or extrapolated DI parameters. In this case, $A$, $B$, $D,$ and $E$ are calculated by applying linear interpolation or extrapolation of the DI parameters derived from the fitting of experimental or theoretical data along the same shell and isoelectronic sequence. The parameter $C$ is always calculated using Eq. \ref{eqn:bethe}.

\section{Ionization cross-sections}

The detailed discussion of the available data used for fitting the cross-sections can be found in the following subsections. In general, we follow the recommendations of AR and D07 in
selecting the most reliable data sets, but also other reviews like \cite{Kallman2007} have been take into account.   We do not repeat their
arguments here, therefore only the relevant differences in the selection criteria and application in the code have been highlighted . Moreover, the multi-searching platform GENIE\footnote{https://www-amdis.iaea.org/GENIE/} has additionally been  used as cross-check. The references for the cross-section data sets used for each isoelectronic sequence (experimental data $e$ or theoretical calculation $t$) are listed in Appendix A.

\begin{figure}
\includegraphics[width=0.55\textwidth]{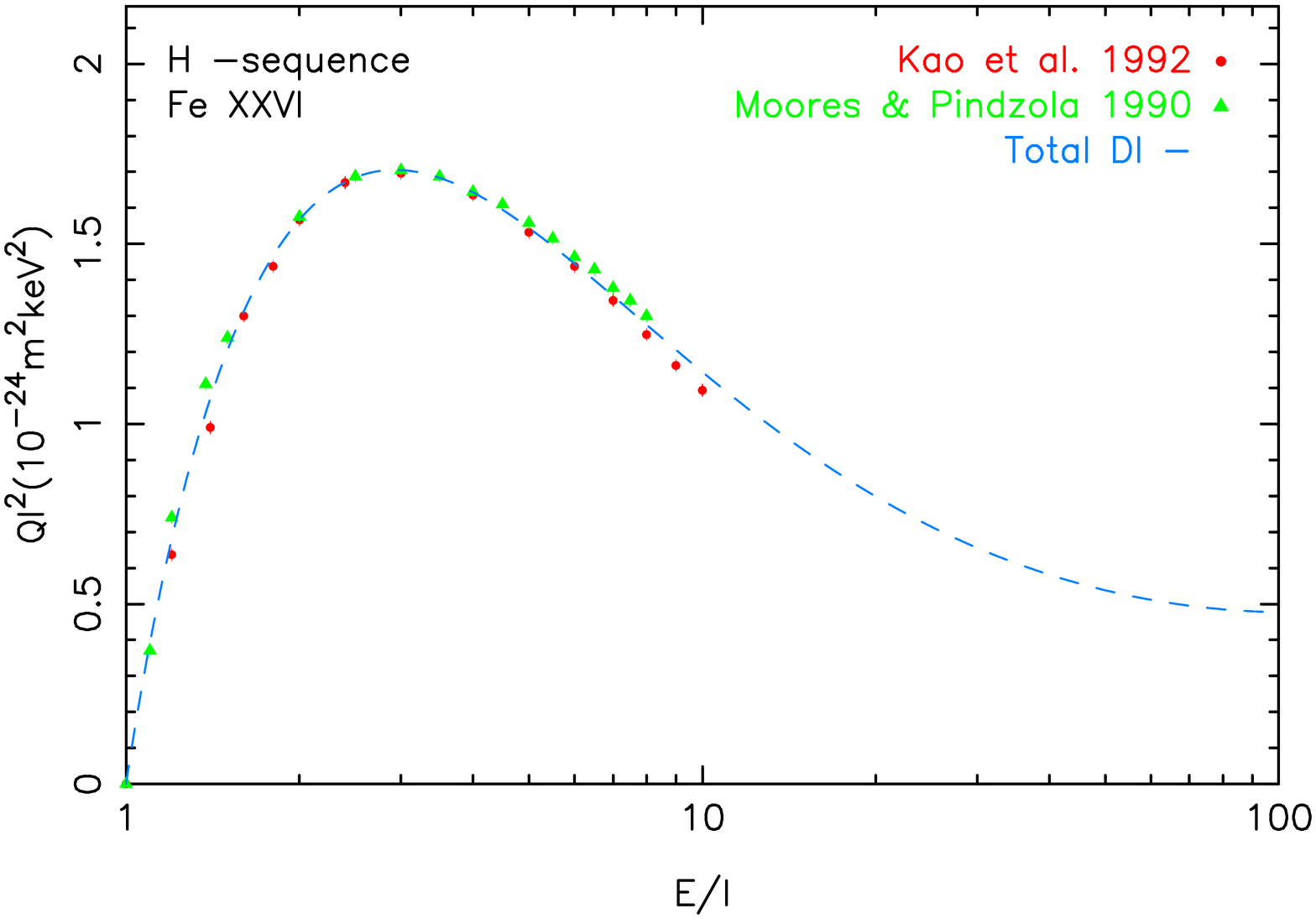}
\caption{Total DI normalized cross-section for \ion{Fe}{XXVI} (dashed blue line) and the measurements of \citet{Kao1992} (red dots) and \citet{Moores1990} (green triangles). Note the presence of relativistic effects for high energies.}
\label{fig:Fig2}
\end{figure}

\subsection{H isoelectronic sequence}

The cross-sections for this sequence include only the direct ionization process from the 1s shell.
For \ion{He}{II}, the cross-sections of \cite{Peart1969} have been selected instead of \cite{Dolder1961}, \cite{Defrance1987} and \citet{Achenbach1984}, because they have a larger extension to the highest energies and an acceptable uncertainty of 12\%, compared to \cite{Dolder1961} with 25\%.

Relativistic effects are important for the high $Z$ elements of this sequence. This is the reason why  the relativistic cross-sections of \cite{Kao1992} and \cite{Fontes1999} for \ion{Ne}{X}; \cite{Kao1992} and \cite{Moores1990} for \ion{Fe}{XXVI}; and \cite{Moores1990} for \ion{Cu}{XXIX} have been chosen. They are, in general, around 5--10\% larger than the non-relativistic ones. These effects are mainly present for high $Z$ elements of the H and He isoelectronic sequences, as can be seen in Fig.~\ref{fig:Fig2}, where the total DI normalized cross-section of \ion{Fe}{XXVI} is shown. For this ion the cross-section increases asymptotically with the energy beyond $u=100$. The measurements of \cite{ORourke2001} for \ion{Fe}{XXVI} has been neglected because they present a considerable experimental error and are in poor agreement with the selected calculations.

The cross-section comparison of this sequence with D07 shows a good agreement for all the elements except for \ion{Be}{IV}. We calculate by linear interpolation in $1/Z$ between \ion{Li}{III} and \ion{B}{V}, which are fitted by experimental measurements. The value at the peak for our interpolation is 20\% lower than the values used by D07. However, it follows a smooth increase, which is consistent with the trend of the rest of the elements in this sequence.

\subsection{He isoelectronic sequence}

\begin{figure}
\includegraphics[width=0.55\textwidth]{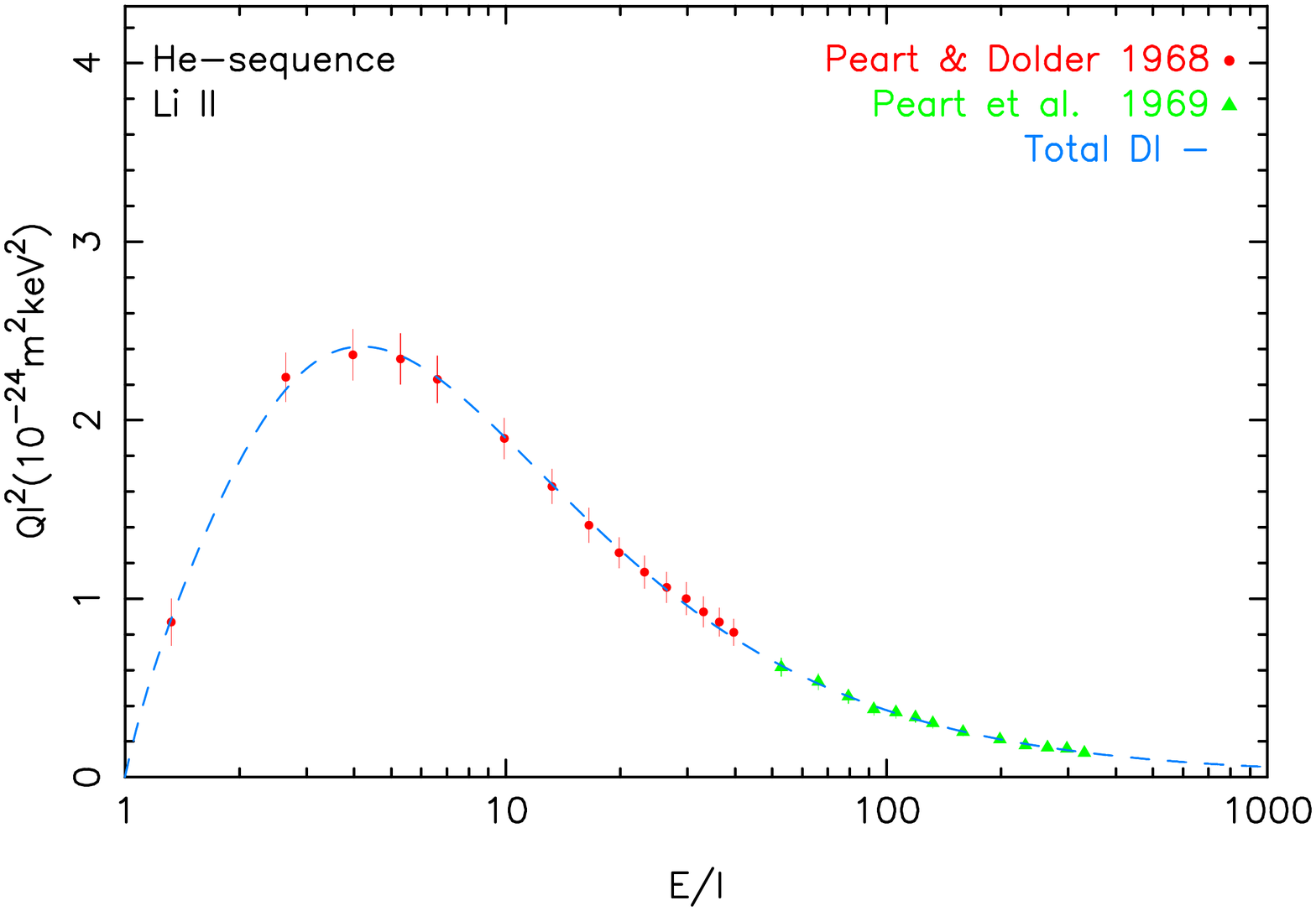}
\caption{Total DI normalized cross-section for \ion{Li}{II} (dashed blue line) and the measurements of \cite{Peart1968} (red dots) and \cite{Peart1969} (green triangles) with their respective experimental error.}
\label{fig:Fig3}
\end{figure}

The He-like ions have an 1s$^2$ structure and the DI process includes ionization of the 1s shell. For \ion{He}{I}, the experimental data of \cite{Shah1988} and \cite{Montague1984b} have been used together with the more recent measurements of \cite{Rejoub2002}.  These data sets are in very good agreement with the cross-sections presented in \cite{Rapp1965}, although the value at the peak is 6\% lower. The final fit has an uncertainty less than 6\%. For \ion{Li}{II}, the measurements of \cite{Peart1968} and \cite{Peart1969} were selected, as shown in Fig.~\ref{fig:Fig3}, an example of the 1s shell DI fitting. For \ion{Be}{III} the same difference with D07 as described for the H-like sequence occurs as well. The peak value using our linear interpolation in $1/(Z-1)$ is 30\% lower than D07.

Since the data range for \ion{Ne}{IX,} as measured by \cite{Duponchelle1997}, is
rather limited, we have supplemented their data by adding the cross-sections at $u=100,$ as interpolated from the calculations of \cite{Zhang1990} for
\ion{O}{VII} and \ion{Fe}{XXV}. 

The relativistic calculations for \ion{O}{VII}, \ion{Fe}{XXV,} and \ion{Zn}{XXIX} of \cite{Zhang1990} yield cross-sections that are about 15\% larger at the higher energies than the corresponding cross-sections interpolated from the theoretical results for N, Na and Fe \citep{Younger1981a,Younger1982a}. This is similar to what
we find for the H-like sequence, and is in agreement with the relativistic effects expected for high $Z$ elements.

\subsection{Li isoelectronic sequence}

Li sequence ions have a structure of 1s$^2$2s and can experience DI in both the 1s and 2s shell with a significant presence of an EA contribution in the outer shell, mainly for highly ionized elements. The DI and EA cross-sections are calculated with the equations described in Section 2.1 and 2.2, respectively. 

\subsubsection{DI: 1s cross-sections}

\cite{Younger1981a} showed that, except for the lower end of the
sequence, the cross-sections are similar to the values for the
He-sequence.  This is confirmed by the work of \cite{Zhang1990}
for O, Fe, Zn, and U.  Wherever needed, we have corrected for a difference in the Bethe constant between the He-like and Li-like sequence using Eq. \ref{eqn:interpar2}.  We note that, again for Ni, the cross-sections of \cite{Pindzola1991} are 15--25\% lower than those of \citet{Zhang1990}.  

In summary, we  used \cite{Younger1981a} for \ion{Li}{I}, \ion{Be}{II}; the corresponding  cross-section of
the He-sequence for \ion{B}{III}, \ion{C}{IV}, \ion{N}{V}; \cite{Zhang1990} for \ion{O}{VI}, \ion{Fe}{XXIV} and
\ion{Zn}{XXVIII}; and linear interpolation in $1/(Z-1)$ for the remaining cross-sections.

\subsubsection{DI: 2s cross-sections}

For \ion{Li}{I}, we follow the recommendations of \cite{McGuire1997} and we fit the convergent close-coupling calculations of \cite{Bray1995} below 70~eV, together with the measurements of \cite{Jalin1973} above 100~eV.

For \ion{B}{III}, \ion{C}{IV}, \ion{N}{V}, \ion{O}{VI,} and \ion{F}{VII,}   high resolution measurements
exist near the EA threshold \citep{Hofmann1990}. These measurements are
systematically higher than the measurements of 
\cite{Crandall1979} and \cite{Crandall1986}, ranging from 9\% for \ion{B}{III}, 24\% for
\ion{N}{V} to 31\% for \ion{O}{VI}.  Moreover, for \ion{C}{IV} Hofmann's data above the EA onset are
inconsistent in shape with both Crandall's measurements and the calculations of
\cite{Reed1992}. Therefore we did not use the measurements of \cite{Hofmann1990}.

For \ion{N}{V}, the measurements of \cite{Crandall1979} are
$\sim$10\% smaller than \cite{Defrance1990}
below 300~eV, but 10\% larger above 1000~eV.  In the intermediate range, where
the EA onset occurs, the agreement is better than 5\%.  
 We  used \cite{Crandall1979}, together with the high energy data of \cite{Donets1981}.

For \ion{O}{VI}, the measurements of \cite{Crandall1986} are
$\sim$10\% smaller than those of \cite{Defrance1990}
below 450~eV and above 800~eV. In the intermediate range, where the EA onset occurs the agreement is good.  We have used both data sets in our fit, but scaled the measurements of \citeauthor{Defrance1990} by a factor of 0.95, and also we have included the high
energy data of \cite{Donets1981}.  Using the statistical
errors in the data sets, the relative weights used in the fit are approximately 5:1:2 for \citeauthor{Crandall1979}, \citeauthor{Defrance1990}, and \citeauthor{Donets1981},
respectively.

For \ion{Fe}{XXIV} and \ion{Zn}{XXVIII,} we  again considered  the relativistic calculations of  
\cite{Zhang1990}.  Their scaled cross-sections for \ion{O}{VI}, \ion{Fe}{XXIV,} and \ion{Zn}{XXVIII}  are
not too different; therefore we interpolate linearly in $1/(Z-1)$ all elements between Ne and Fe, and similarly between Fe and Zn.

For Ti to Fe,  measurements also exist at about 2.3 times the ionization threshold \citep{Wong1993}  with an uncertainty of 10\%, which are also proposed by D07. The ratio of these observed cross-sections to the calculations are 0.83, 0.81, 0.85, 0.84, and 0.97, respectively for $Z=$ 22--26. Given the measurement uncertainty, and the agreement of the calculations of \citeauthor{Zhang1990} in the region of overlap with those of \cite{Chen1992}, we  finally decided to use  the calculations of \citeauthor{Zhang1990}.

As for the 1s cross-sections in the H, He, and Li sequences, the cross-sections of \cite{Younger1982a} for \ion{Fe}{XXIV} are 5\% smaller than those of \citeauthor{Zhang1990} at the highest energies, instead of the typically 15\% for the 1s
cross-sections.  Thus relativistic effects are slightly less important, which can be understood given the lower ionization potential for the 2s shell, compared to the 1s shell.  

\begin{figure}
\includegraphics[width=0.55\textwidth]{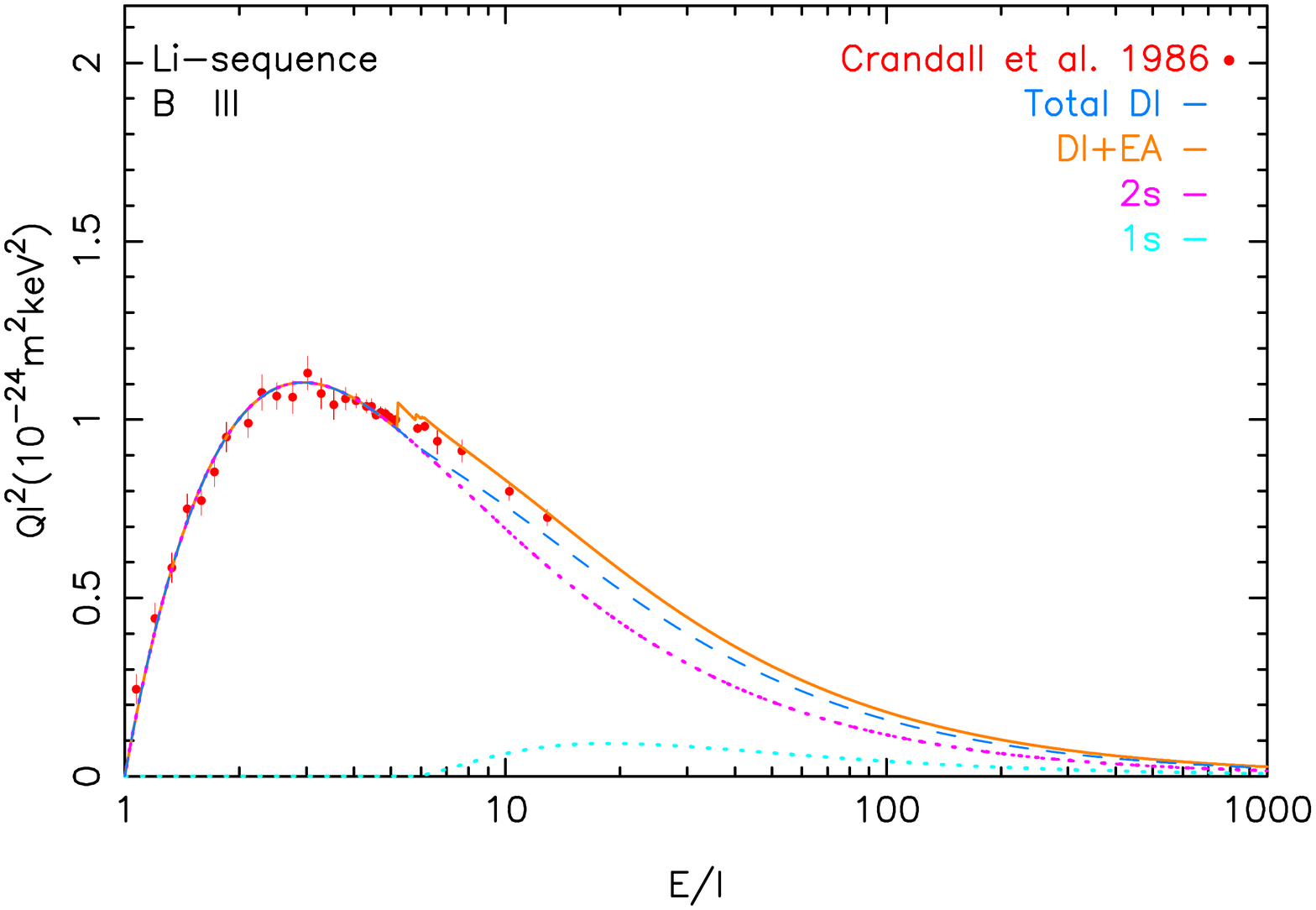}
\caption{Total DI (dashed blue line) and DI plus EA (orange line) normalized cross-section for \ion{B}{III}, where the DI contribution of the 1s shell is shown as the dotted cyan line and the 2s shell in magenta. The measurements of \cite{Crandall1986} (red dots) with the  experimental uncertainties are also included.} 
\label{fig:Fig4}
\end{figure}

\subsubsection{EA contribution}

Fits to the calculations of \cite{Sampson1981} were used to approximate the shape of the EA contribution.  The contributions, which are due to excitation towards $n$=2, 3, 4, and 5, are treated separately.  A comparison of the results of  \citeauthor{Sampson1981} $(Q_{\rm SG}$) with the more sophisticated
calculations of \cite{Reed1992} and \cite{Chen1992}
($Q_{\rm RC}$) just above the $1s-2p$ excitation threshold for $Z$= 6, 9, 18, 26,
and 36 gives for the ratio $Q_{\rm RC}/Q_{\rm SG}$ values of 0.52, 0.64, 0.77,
1.18, and 2.02, respectively.  The following approximation has been made to these
data:
\begin{equation} 
\label{eqn:eali_scale}
Q_{\rm RC} = [0.54 + 1.33\times 10^{-4}~Z^{2.6}] Q_{\rm SG}.
\end{equation} 
A similar tendency is noted by AR.  A more detailed analysis shows that for
larger energies the discrepancy is slightly smaller.  Unfortunately, \citeauthor{Reed1992} and \citeauthor{Chen1992} only give  the EA cross-section near the excitation thresholds.
Therefore, we decided to retain the calculations of 
\cite{Sampson1981}, but to scale all EA cross-sections using
Eq. \ref{eqn:eali_scale}.  We note that, for this isoelectronic sequence, the EA
contribution is, in general, smaller than $\sim$10\%, and thus slight
uncertainties in the EA cross-section are not very great in the total
ionization cross-section. Figure~\ref{fig:Fig4} shows an example of Li-like ion cross-sections scaling.

\subsection{Be isoelectronic sequence}

The Be sequence elements have a structure of 1s$^2$2s$^2$ and can experience DI through collisions in the 1s and 2s shells. There is also an EA contribution. Moreover, in experimental data, some elements, like \ion{C}{III}, \ion{N}{VI,} and \ion{O}{V} often show a high population of ions in metastable levels 1s$^2$2s2p. 

\subsubsection{DI: 1s cross-sections}

\begin{figure}
\includegraphics[width=0.55\textwidth]{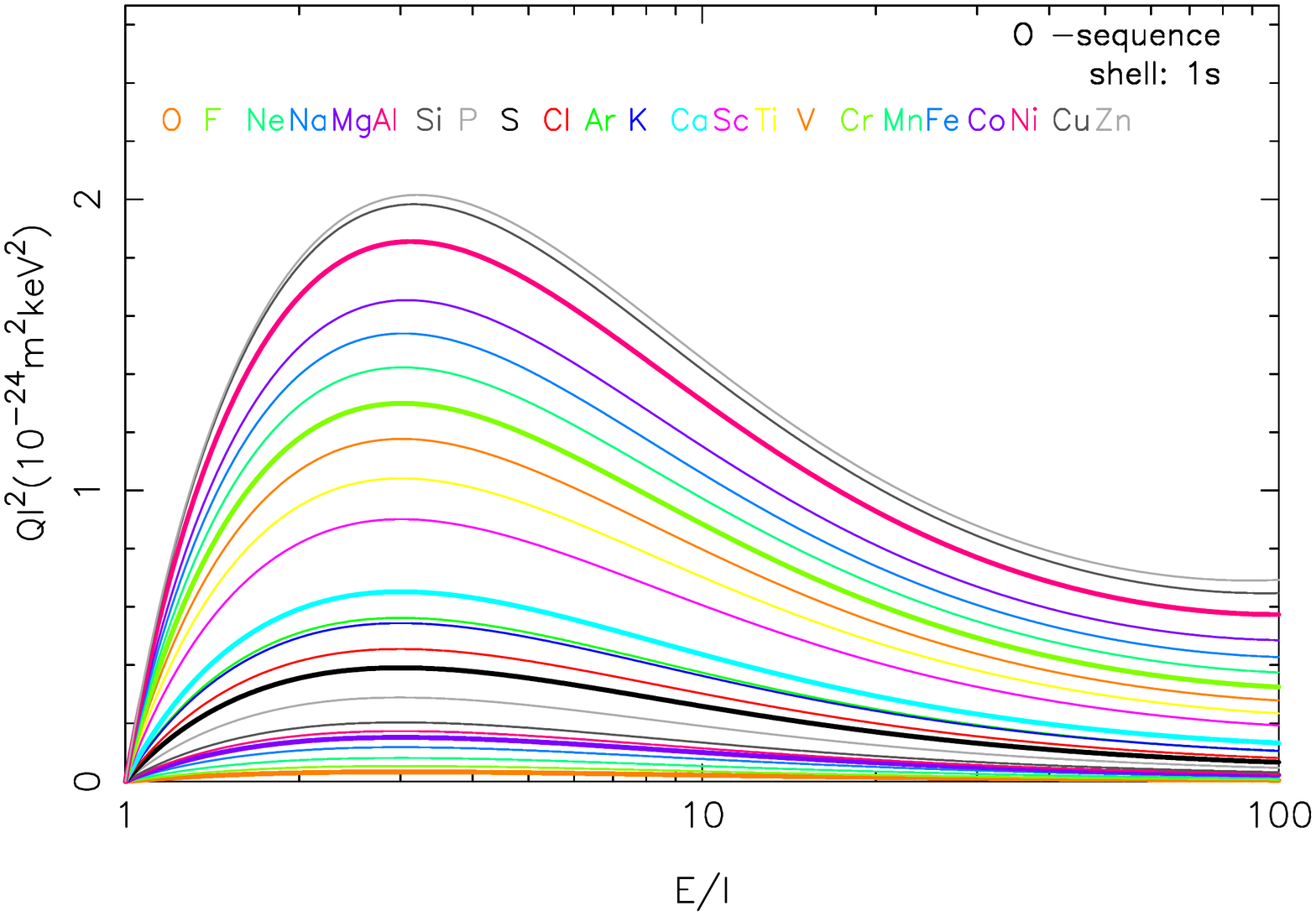}
\caption{Total DI cross-section for the 1s subshell for all elements of the oxygen sequence using interpolation with equation (\ref{eqn:interpar2}).} 
\label{fig:Fig5}
\end{figure}

The 1s cross-sections for all elements in the Be to Zn isoelectronic sequences have been calculated with Eq.\ref{eqn:interpar2} and, using as a reference, the parameters obtained for the 1s inner shell of Li-like ions. An example can be seen in Fig.~\ref{fig:Fig5} for the oxygen isoelectronic sequence. 

A comparison of some K-shell measurements compiled by \cite{Llovet2014}, for example for \ion{C}{I}, \ion{Al}{I,} and \ion{Ti}{I} \citep{Limandri2012},  demonstrated a good agreement with the maximum difference between the measurements and the calculations with (\ref{eqn:interpar2}) at the peak for \ion{Ti}{I} of less than 15\%.

\subsubsection{DI: 2s cross-sections}

The measurements in this sequence are often greatly affected  by metastable ions (see the discussion in AR). As mentioned in \cite{Loch2003} and \cite{Loch2005}, it is essential to know the ratio of  the metastable configuration for an accurate determination of the ground-state cross-section.

D07 proposes  using the measurements of \cite{Falk1983a} for \ion{B}{II}, which we discard owinge to the existence of a significant population of ions in metastable levels, which results in a ground-state cross-section higher than that proposed by \cite{Fogle2008} for \ion{C}{III}, \ion{N}{IV,} and \ion{O}{V}. Fogle's measurements use the crossed-beam apparatus at Oak Ridge National Laboratory. In this experiment, it was possible to measure the metastable ion fractions present in the ion beams in the 1s$^2$2s2p levels, which were used to infer the rate coefficients for the electron-impact single ionization from the ground state and metastable term of each ion. Considering these mentioned rates in the paper for the ground cross-sections calculations, they are in good agreement (error of $\sim$7\% for \ion{C}{III} and $\sim$2\% for \ion{N}{IV} and \ion{O}{V}) with the cross-sections obtained by the \cite{Younger1981d} theoretical calculation. The measurements of \cite{Loch2003} for \ion{O}{V} have been neglected because it was not possible to determine the metastable fraction at the experimental crossed-beam.

The measurements
of \citeauthor{Bannister1996a} for \ion{Ne}{VII} are consistent with the \cite{Duponchelle1997} ones at high energies, but show a bump around 280~eV, and are finally rejected. For \ion{S}{XIII}, \cite{Hahn2012a}  eliminated all metastable levels using hyperfine induced decays, combined with an ion storage ring, obtaining a total cross-section with $1\sigma$ uncertainty of 15\%. The measurements are in very good agreement with the theoretical data of \cite{Younger1981c} and distorted-wave calculation of D07.

Lacking more reliable measurements for this isoelectronic sequence, and given the reasonable agreement with the measurements for \ion{Ne}{VII}, we base our cross-sections on the theoretical calculations of Younger \citep{Younger1981d,Younger1982a} for \ion{F}{VI}, \ion{Ar}{XV,} and \ion{Fe}{XXIII}. The calculations of \ion{Fe}{XXIII} have been multiplied by a scaling factor of 1.05, to account for these effects, as are present in the Li-sequence for 2s electrons.

\subsubsection{EA contribution}

For all ions of this sequence, we also include the EA contribution according to \cite{Sampson1981}, although the contribution is small (in general smaller
than $\sim$5\%).  A comparison of the results of \citeauthor{Sampson1981} $(Q_{\rm
SG}$) with the more recent calculations of \cite{Badnell1993} ($Q_{\rm
BP}$), which include calculations  for only Fe, Kr, and Xe just above the 1s-2p excitation threshold, was performed. This  shows a
systematic trend that can be approximated by
\begin{equation}
\label{eqn:ea_be}
Q_{\rm BP} = [0.70 + 1.46.10^{-3}~Z^2]Q_{\rm SG}.
\end{equation}
We assume that the rest of the elements of this isoelectronic sequence present the same behaviour. Therefore, we use the calculations of Sampson \& Golden (1981), but scale all EA
cross-sections to the results of \citeauthor{Badnell1993} using (\ref{eqn:ea_be}).

\subsection{The B isoelectronic sequence}

The elements of the B-like sequence (1s$^2$2s$^2$2p) have an EA contribution in the outer shell that is relatively small \citep{Yamada1989a,Duponchelle1997,Loch2003}.

\subsubsection{DI: 2s cross-sections}

\cite{Younger1982a} shows that, for the iron ions of the Be to Ne 
sequences, the 2s cross-section is approximately a linear function of the number of the 2p electrons present in the ion.  Following AR, we assume such a linear
dependence to hold for all ions of these sequences.  Thus, from the 2s cross-sections for the Be-sequence and those of the Ne-sequence, the 2s cross-sections for all ions between Na--Zn for intermediate sequences (B-like, C-like, N-like, O-like, and F-like) are obtained by
linear interpolation plus the Bethe coefficient difference correction applying  Eq. \ref{eqn:interpar}.

For ions of B to F in the B-F isoelectronic sequences, we cannot use the above interpolation since,
in this case, there are no ions in the Ne-sequence. AR assume that the
2s cross-section of the Ne-sequence minus the 2s cross-section of the
Be-sequence depends linearly upon the atomic number $Z$; since our
procedure is slightly different from \citeauthor{Arnaud1985}, we cannot confirm clear linear trends in our data.
For that reason, we use for the ions from B to F a linear extrapolation of
the difference coefficients given by AR:
\begin{eqnarray} 
\label{eqn:b_ne}
A({\rm Ne\ seq., 2s}) - A({\rm Be\ seq.,2s}) &= 4.20 - 0.1658~Z,
&  \nonumber\\
B({\rm Ne\ seq., 2s}) - B({\rm Be\ seq.,2s}) &= -0.42 - 0.1313~Z,
& \nonumber \\
C({\rm Ne\ seq., 2s}) - C({\rm Be\ seq.,2s}) &= -0.05 + 0.0088~Z,
& \nonumber \\
E({\rm Ne\ seq., 2s}) - E({\rm Be\ seq.,2s}) &=  -18.87 + 0.8240~Z,
&  
\end{eqnarray}
where $Z$ is the atomic number and $A$, $B,$ and $E$ are in units of
$10^{-24}$~m$^2$keV$^2$.

\subsubsection{DI: 2p cross-sections}

For \ion{B}{I,} we  included the CHIANTI data obtained from \cite{Tawara2002} (D07). The data of \cite{Aitken1971} for \ion{C}{II} are slightly higher than the measurements of \cite{Yamada1989a}, especially near the threshold. Nevertheless, we use both data sets in our fit, with a larger weight given to the data of \citeauthor{Yamada1989a}

For \ion{N}{III}, we  chose \cite{Aitken1971} and \cite{Bannister1996b} proposed by D07 because both data sets extend from near threshold to $u=20$ and, besides, they are in relatively good agreement, except below the peak where the data of \cite{Bannister1996b} are $\sim$5\% higher. 

The most recent measurement for the B-sequence is that of \cite{Hahn2010} for \ion{Mg}{VIII}. The innovative aspect of the \citeauthor{Hahn2010} data is the use of an ion storage ring (TSR) for the measurements. This new experimental technique achieves a radiative relaxation of ions to the ground state after being previously stored  long enough in the TSR, decreasing considerably the contribution of possible metastable ions. The data show a 15\% systematic uncertainty owing to the ion current measurement. Nevertheless, the data are in good agreement with the distorted-wave calculations with the GIPPER  \citep{Magee1995} and FAC \citep{Gu2002} codes, within the experimental uncertainties.

The theoretical data for \ion{Fe}{XXII} are based upon \cite{Zhang1990} for Ne-like iron. Following \cite{Younger1982a}, we assume that the
scaled 2p cross-section for B-like to Ne-like iron is a linear function
of the number of 2p-electrons; we account for the slight difference in
2p$_{1/2}$ and 2p$_{3/2}$ cross-sections in the work of \citeauthor{Zhang1990}. Finally, we use their data for Se ($Z$=34) to
interpolate the ions between Fe and Zn on these sequences.

\subsubsection{EA contribution}

For all ions of this sequence, we include the EA contribution according
to \cite{Sampson1981}, although the contribution is small (in
general less than $\sim$2.5\%).  A comparison of the results of
\citeauthor{Sampson1981} $(Q_{\rm SG}$) with the calculations of
\cite{Badnell1993} ($Q_{\rm BP}$) for Fe, Kr and Xe just above
the 1s-2p excitation threshold, shows a systematic trend that can be approximated by
 \begin{equation}
 \label{eqn:ea_b}
Q_{\rm BP} = [0.92 + 7.45\times 10^{-5}~Z^2]Q_{\rm SG}.
\end{equation} 
We retain the calculations of \citeauthor{Sampson1981}, but scaled all EAcross-sections to the results of \citeauthor{Badnell1993} using Eq. \ref{eqn:ea_b}.

\subsection{C isoelectronic sequence}

The ions of the carbon isoelectronic sequence (1s$^2$2s$^2$2p$^2$) can be directly ionized by the collision of a free electron with electrons in the 1s, 2s, and 2p shells; the same holds for all sequences up to the Ne-like sequence. There is no evidence for a significant EA processes in the C to Ne sequences.

\begin{figure}
\includegraphics[width=0.55\textwidth]{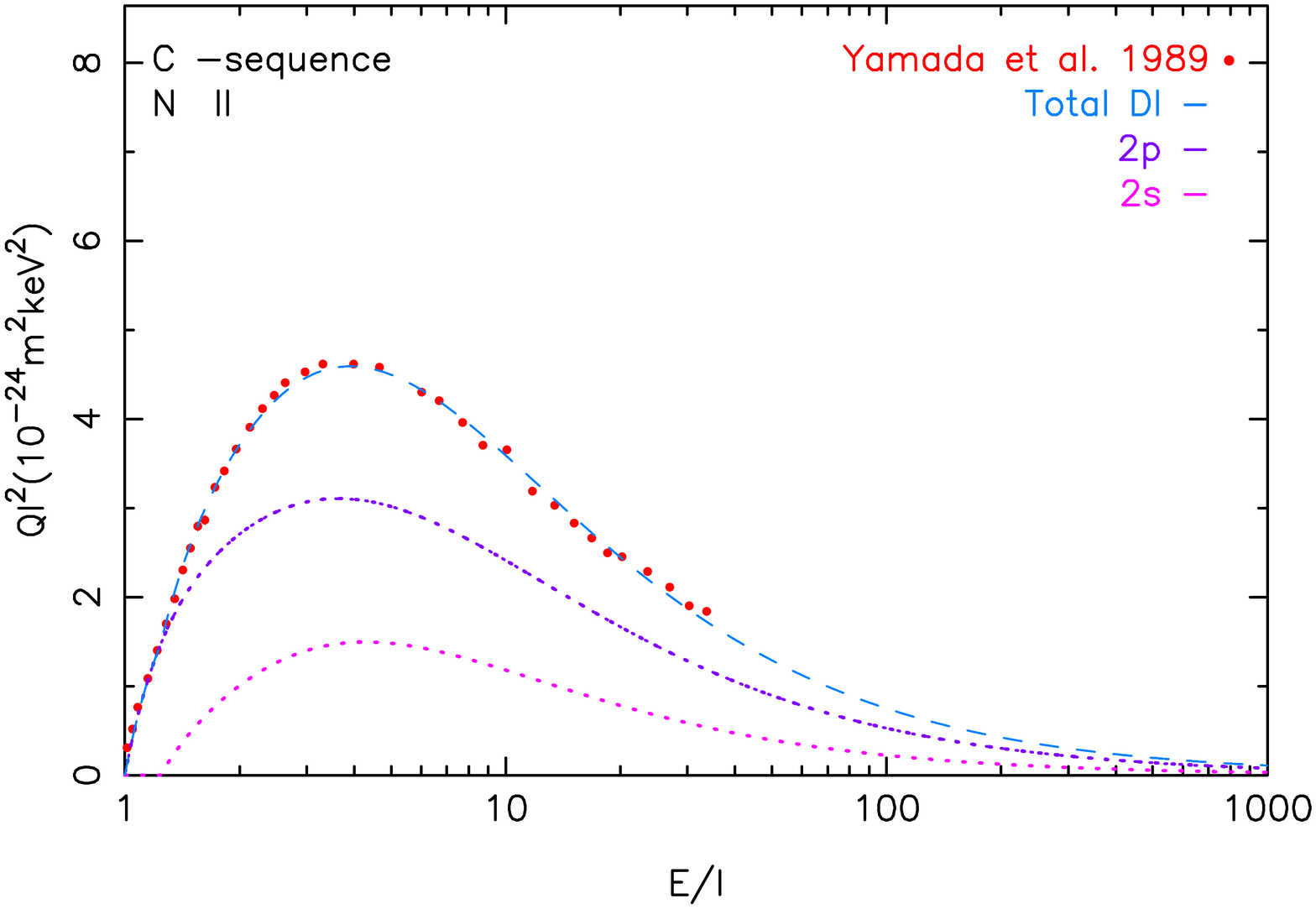}
\caption{Total DI normalized cross-section for \ion{Ne}{II} (dashed blue line) and the measurements of \cite{Yamada1989a} (red dots) with the experimental uncertainties.}
\label{fig:Fig6}
\end{figure}

\subsubsection{DI: 2p cross-section}

For \ion{O}{III}, we use the measurements of \cite{Aitken1971}, \cite{Donets1981}, and \cite{Falk1980}. The first two are provided up to ten times the threshold.  The \cite{Aitken1971} measurements are $\sim$15\% lower than those of \cite{Falk1980} beyond the cross-section peak. We also use the data of \cite{Donets1981} for the high energy range. Figure~\ref{fig:Fig6} shows an example of the DI contribution for the 2s and 2p shells.

\subsection{N isoelectronic sequence}

\subsubsection{DI: 2p cross-section}

The measurements for \ion{Si}{VIII} \citep{Zeijlmans1993} were correctly fitted using Eq. \ref{eqn:younger_ext},  obtaining a maximum uncertainty of $\sim$6\%. The peak value of \ion{Si}{VIII} compared with CHIANTI data is $\sim$10\% lower. This difference also affects the cross-sections of interpolated components between \ion{Ne}{IV} and  \ion{Si}{VIII}.

The data of \cite{Yamada1989a} for \ion{O}{II} are about 5\% higher than
the older data of \cite{Aitken1971}; our fit lies between both
sets of measurements. For \ion{O}{II} and \ion{Ne}{IV}, the high energy measurements of \cite{Donets1981} are significantly higher than our fit, including that data
set; we have therefore discarded these measurements for these ions.

\subsection{O isoelectronic sequence}

\subsubsection{DI: 2p cross section}

 We note that the  measurements for \ion{Si}{VII} \citep{Zeijlmans1993} could be affected by metastable ions that show an increase between 10-20\% of the cross-section at the peak compared with the distorted-wave calculations. This is the reason we neglect this data set. For \ion{Ne}{III}, we discard the high energy measurements of \citeauthor{Donets1981} since these are 30\% below our fit including those data.

\subsection{F isoelectronic sequence}

\begin{figure}
\centering
\includegraphics[width=0.55\textwidth]{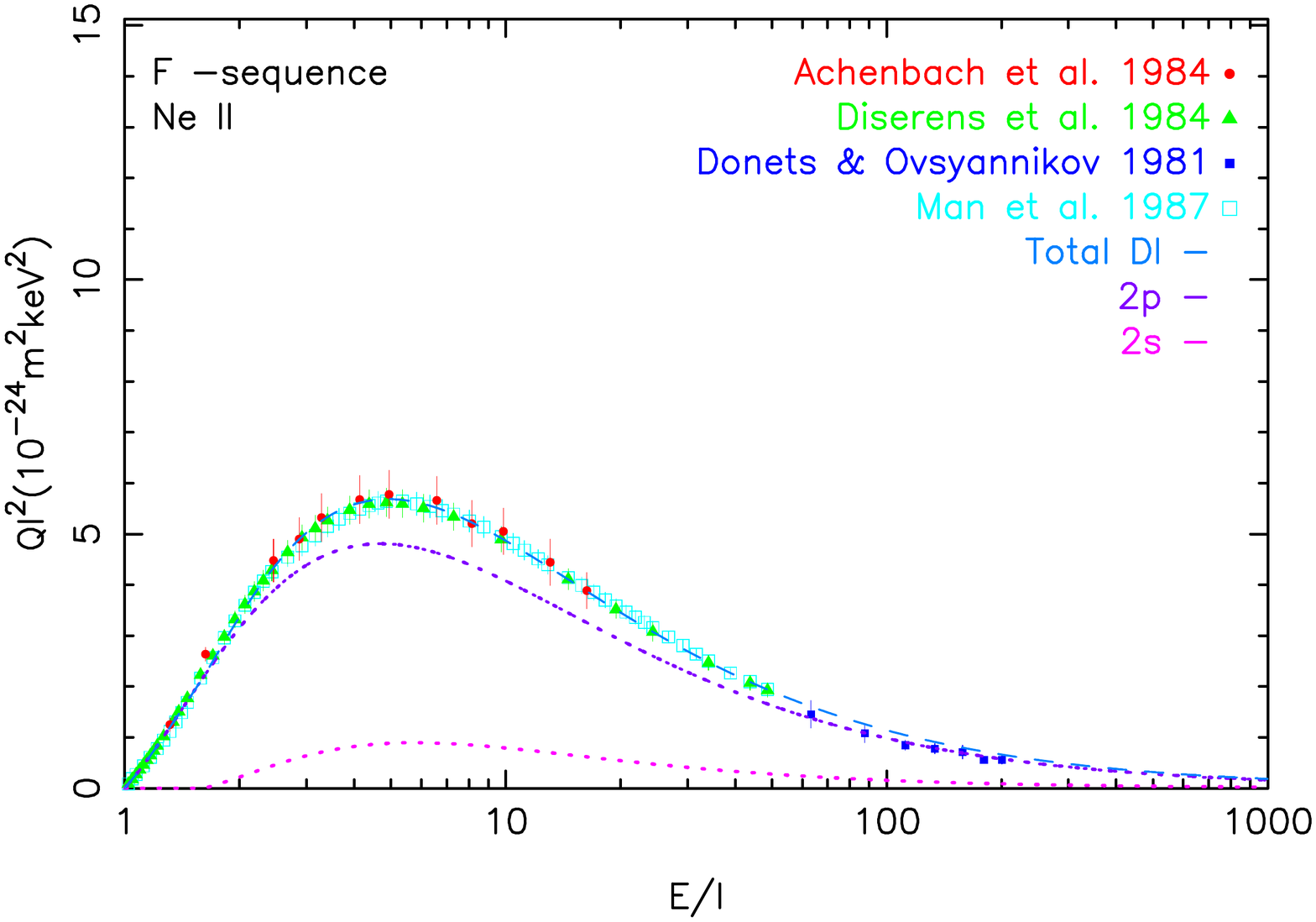}
\caption{The DI normalized cross-section for \ion{Ne}{II} (dashed blue line) and the measurements of \cite{Achenbach1984} (red dots), \cite{Diserens1984} (green triangle), \cite{Donets1981}  (blue square) and   \cite{Man1987b} (cyan square) with their respective experimental uncertainties.}
\label{fig:Fig7}
\end{figure}

\subsubsection{DI: 2p cross-section}

The \cite{Yamada1989a} measurements were included for \ion{F}{I} up to $u=10$. The most recent measurements in this sequence are from \cite{Hahn2013} for \ion{Fe}{XVIII} up to energies of $u=3$. The measurements given by \cite{Hahn2013} are 30\% lower than the values provided by \cite{Arnaud1992} and 20\% lower than D07. This is achieved by the new experimental technique of the ion storage ring. We combine these data with the theoretical calculations of \cite{Zhang1990} for high energies. These theoretical data were obtained directly from the total cross-section modelling for the 2p shell, as explained in Section 3.5.2. The Fig.~\ref{fig:Fig7} above shows the DI fitting of four different experimental measurements for \ion{Ne}{II}.  

\subsection{Ne isoelectronic sequence}

\subsubsection{DI: 2s cross-section}
 
 For Na, Mg and Al \cite{Younger1981c}  calculates the 2s cross-sections in the Na-like sequence. For the high $Z$ end of the sequence
(Ar and Fe), the difference between the Ne-like and Na-like 2s cross-section is, in general, at most a few percent. Accordingly, we assume that,
for the low $Z$ end of the Ne-sequence, the shape of the cross-section is at
least similar to that of the Na-sequence. 

Therefore, we have extrapolated the Na-like data of \cite{Younger1981c} to obtain the cross-sections of  \ion{Na}{II}, \ion{Mg}{III}, \ion{Al}{IV}, \ion{P}{VI,} and  \ion{Ar}{IX}.  We  found that the ratio of the Ne-like to the Na-like 2s cross-section is about 1.38, 1.23, 1.06, and 1.00 for the elements Na, Mg, Al, and Ar. We  included a scaling factor of 1.00 for P. Our adopted  2s ionization cross-section for the Ne-like ions \ion{Na}{II}, \ion{Mg}{III}, \ion{Al}{IV}, \ion{P}{VI,} and  \ion{Ar}{IX} are thus based upon the corresponding Na-like cross-section, multiplied by the above scaling factors. Lacking other data, for \ion{Ne}{I} we simply used the correlation Eq. \ref{eqn:b_ne} between Ne-like and Be-like 2s cross-sections.
For the remaining elements from Si and higher, we use linear
interpolation in $1/(Z-3)$.

\subsubsection{DI: 2p cross-sections}

We  used  the calculations of \cite{Younger1981b} for the 2p shell of \ion{Al}{IV} instead of the \cite{Aichele2001} measurements because, as they explain, their data contain a 20\%  contribution from metastable ions contamination.

For \ion{Ar}{IX}, we did not use the data of \cite{Zhang1991}, because they
contain a 3\% contribution of a metastable state, which is strongly
auto-ionising. The contribution of this metastable state, which is
described well by the calculations of \cite{Pindzola1991}, makes the
measured cross-section $\sim$5\% higher at 1~keV; however, owing to the
complex ionization cross-section of this metastable state, we do not
attempt to subtract it, but merely use the data of \cite{Defrance1987} and \cite{Zhang2002}
for this ion, which appears to be free of metastable
contributions.

\begin{figure}
\includegraphics[width=0.55\textwidth]{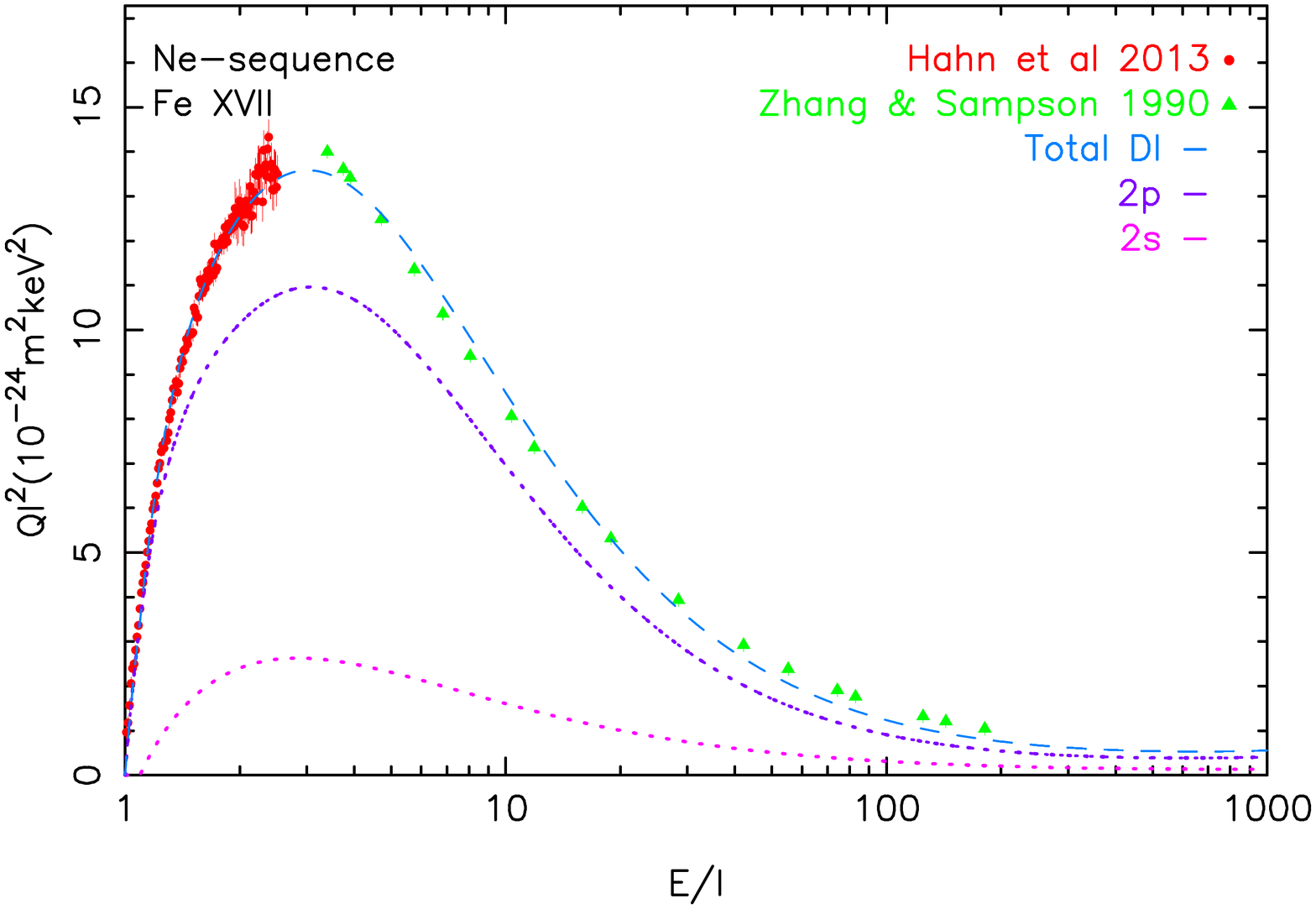}
\caption{Total DI normalized cross-section for \ion{Fe}{XVII} (dashed blue line) and the measurements of \cite{Hahn2013} (red dots), and calculations of \cite{Zhang1990} (green triangles).}
\label{fig:Fig8}
\end{figure}

We have used the measurements of \cite{Hahn2013} for \ion{Fe}{XVII} up to energies close to $u=3$, together with \cite{Zhang1990} for high energies, as in the same case of \ion{Fe}{XVIII}, see Fig.~\ref{fig:Fig8}.

\subsection{Na isoeletronic sequence}

\subsubsection{DI: 2s and 2p cross-section}

We use the theoretical calculations of \cite{Younger1981c} for \ion{Mg}{II}, \ion{Al}{III}, \ion{P}{V,} and \ion{Ar}{VIII} and \cite{Pindzola1991} for \ion{Ni}{XVIII}. The
remaining elements have been interpolated, except for \ion{Na}{I,} for which we  adopted the scaled \ion{Mg}{II} parameters.

\subsubsection{DI: 3s cross-section}

We  decided not to include the measurements for \ion{Ar}{VIII} \citep{Rachafi1991,Zhang2002} and \ion{Cr}{XIV} \citep{Gregory1990}, because they are higher and lower, respectively, compared with the other elements on this sequence. The \ion{Ar}{VIII} measurements are probably affected by the presence of resonant excitation double Auger ionization (REDA).

\ion{Ti}{XII} was  fitted using \cite{Gregory1990}, although there are only measurements up to $u=3$. For this reason we  included some values from  \cite{Griffin1987} calculation for higher energies. The data of \cite{Gregory1990} are about 10\% higher than the calculations of \cite{Griffin1987}, therefore we  decided to apply a scaling factor of 0.9.

For \ion{Fe}{XVI,} the measurements of \cite{Gregory1987} and \cite{Linkemann1995} were used, which extend till $u=2$. The data of \cite{Gregory1987} are 30\% higher than those of \cite{Linkemann1995}, resulting in a fit with values around 15\% higher than proposed by the \cite{Griffin1987}. To achieve a better agreement of 5-10\%, we  applied a scaling factor of 0.9 to the \cite{Gregory1987} measurements.
Finally, we included the theoretical calculations of \cite{Pindzola1991} for \ion{Ni}{XVIII}, which agree with \cite{Griffin1987} better than 10\%.

The total cross-sections obtained with our method are systematically higher than D07 by 10-30\%. For several elements, the DI level seems to be of the same order and the main difference is related to EA contributions. A possible explanation could be that we use the calculations of \cite{Sampson1982}, which include more excitation transitions (from 2s, 2p, and 3s subshells to the $n$s, $n$p and $n$d subshells with $n$=3--5), while D07 use the FAC EA calculation scaled by a certain factor for excitation into 2$^7$ 3$l$3$l'$ and 2$^7$ 3$l$4$l'$.

\subsubsection{EA contribution}

For the low $Z$ elements of this sequence (\ion{Mg}{II}, \ion{Al}{III} and \ion{Si}{IV}) the
theoretical calculations of \cite{Griffin1982} for the EA
contribution fail to correctly model  the measurements, mainly due to
too large 2p$\rightarrow$3p and 2p$\rightarrow$4p cross-sections,
but also becuase of  the presence of REDA contributions in the measured cross-sections \citep{Muller1990,Peart1991b}.  The measured
cross-sections show a distinct EA contribution, but not with the
sharp edges that are usually produced  by theory owing to limitations in the way the EA contribution is calculated. Therefore, we fitted the measured cross-sections of \ion{Mg}{II}, \ion{Al}{III,} and \ion{Si}{IV} to (\ref{eqn:eamewe}) after subtracting the DI contributions.

For \ion{Na}{I} and \ion{Mg}{II,} there are no signs of the EA onset owing to the regularity of the measurements although some REDA contributions could be present. For \ion{Al}{III} and \ion{Si}{IV,} we have followed the same procedure chosen by D07 for scaling all the EA cross-sections to recreate the measured values. Therefore, we  retained the calculations of \cite{Sampson1982}, but scaled by a factor of 0.4 for \ion{Al}{III} and 0.5 for \ion{Si}{IV}. The rest of elements have been maintained with a scaling factor equal to 1.0.

\subsection{Mg isoelectronic sequence}

\subsubsection{DI: 2s and 2p cross-section}

The 2s and 2p cross-sections for all elements from the Mg to Zn isoelectronic sequences have been calculated with Eq. \ref{eqn:interpar} using, as a reference, the parameters obtained for the 2s and 2p subshell, respectively, of the previous isoelectronic sequence. Therefore, for the Mg-sequence, the reference parameters for all elements are taken from the Na-sequence.

\begin{figure}
\includegraphics[width=0.50\textwidth]{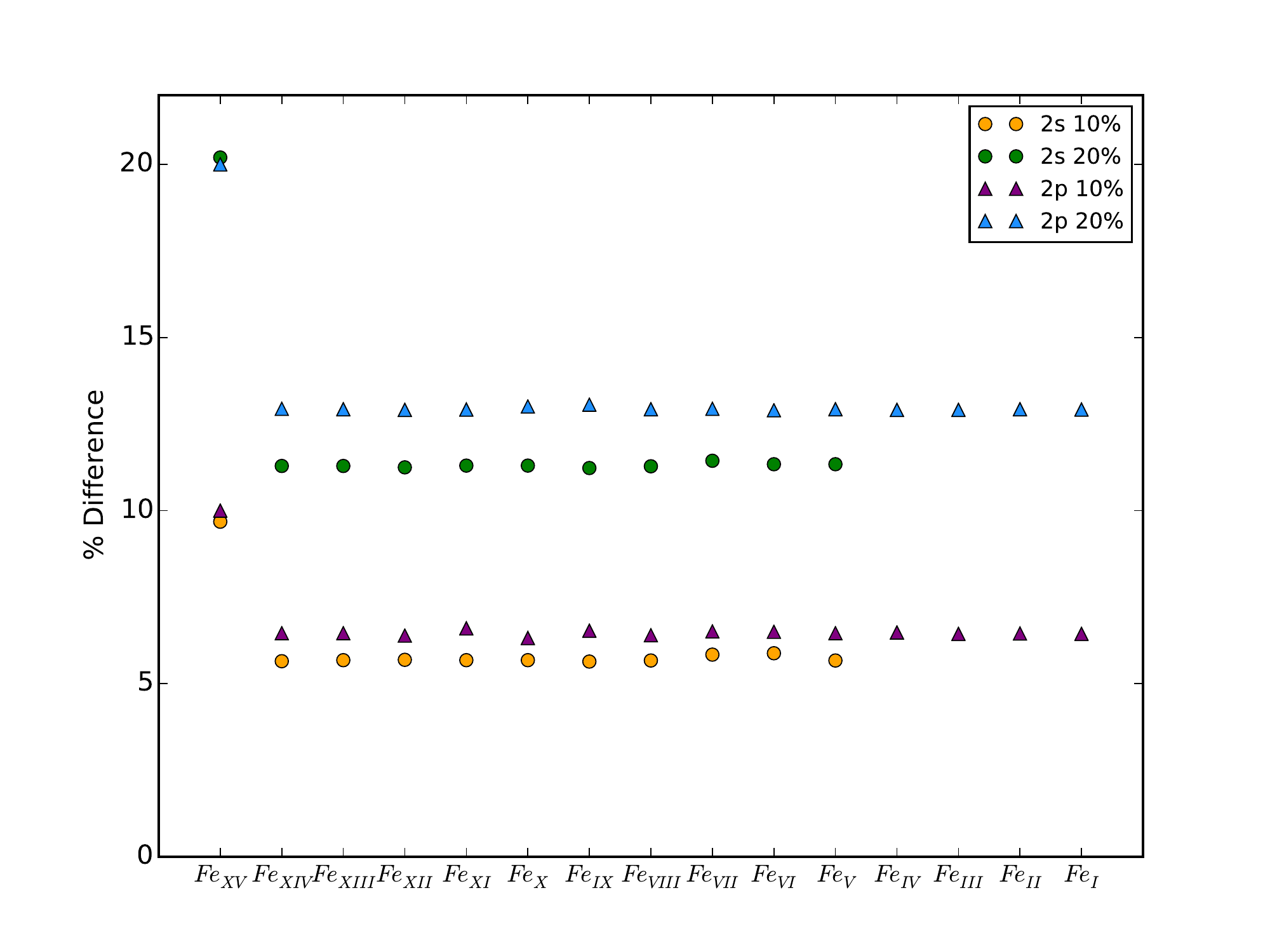}
\caption{Error propagation along Fe ions after applying a 10\% and 20\% increase in the 2s shell (orange and green circles) and the 2p shell (purple and blue triangles) DI parameters of \ion{Fe}{XV}.}
\label{fig:Fig18}
\end{figure}

To evaluate the robustness of this method, we  introduced a 10\% and 20\% increase in the $A$ to $E$ parameters of the \ion{Fe}{XV} 2s shell and analysed how it affects the 2s shells of Fe ions for the following isoelectronic sequences. If we compare the difference in the peak of the cross-sections, from the Al to Ti sequences (there is no  2s shell contribution for the V to Fe-sequence) the error is reduced to 5-6\% and 11\%, respectively. This difference is maintained almost constantly for all the sequences as seen in Fig. \ref{fig:Fig18}. We  also evaluated the impact in the calculation of the outer shell DI cross-sections using the fitting procedure explained in Section 2.3 and the effects are negligible, being the maximum difference of 0.03\%, 0.07\%, and 0.06\%, for an initial increase of 10\%, 20\%, and 50\%, respectively. The main reason for this behaviour is because the 2s shell cross-sections and their contribution to the single ionization is much lower than the outer shells. Therefore, a variation in the DI parameters of the 2s shell has no appreciable effects on the other shell cross-sections.

We  performed the same study for the 2p shell as explained above for the 2s shell. The results are slightly similar and the same conclusions are applicable in this case. 

\begin{figure}
\includegraphics[width=0.55\textwidth]{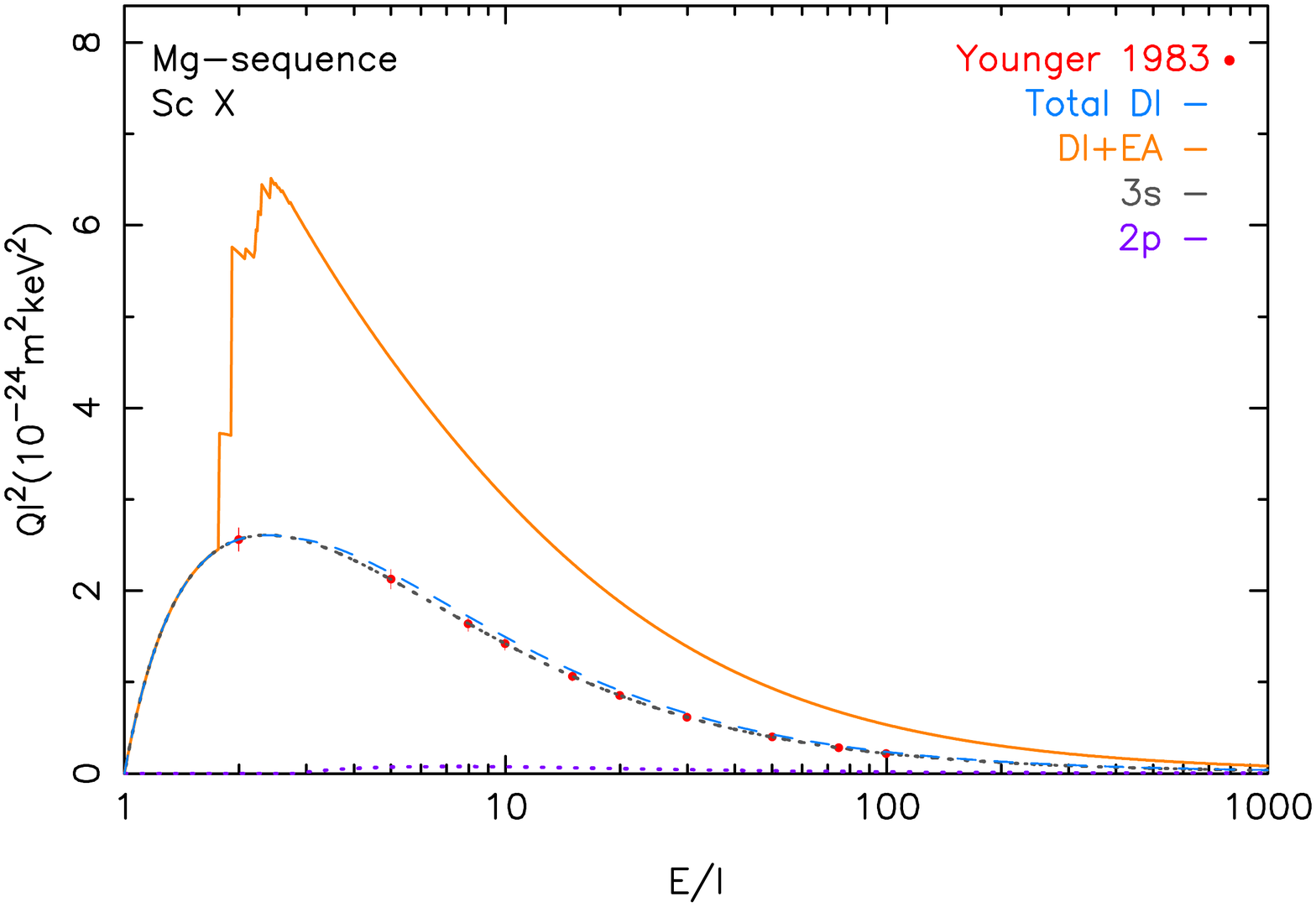}
\caption{Total normalized cross-section for \ion{Sc}{X} (orange line), DI cross-section (dashed blue line), 3s (pointed grey line), 2p (pointed purple line) and the calculations for the 3s shell of \cite{Younger1982b} (red dots).}
\label{fig:Fig10}
\end{figure}

\subsubsection{DI: 3s cross-sections}

Following the discussion in \cite{McGuire1997}, we  scaled-down the
experimental results for Mg~I and Al~II by multiplying by a factor of 0.8.
For \ion{Si}{III} and \ion{S}{V} the measurements of \cite{Djuric1993b} and \cite{Howald1986}, respectively, were  omitted because the measurements present clear evidence of metastable ions.

In the case of \ion{Ar}{VII}, the measurements of \cite{Chung1996} do not show evidence of 3s3p $^3$P metastable ions unlike \cite{Djuric1993b}, \cite{Howald1986} and \cite{Zhang2002}. However, they only extend up to $u=6$, where they seem to be in good agreement with \cite{Zhang2002}, which contains data till 30 times the threshold. For this reason, we  used the \cite{Chung1996} measurement adding the \cite{Zhang2002} cross-sections above $u=5$.

\cite{Bernhardt2014} present recent measurements for \ion{Fe}{XV} in the range 0--2600~eV. \citeauthor{Bernhardt2014} use the TSR storage-ring technique, also applied to several measurements of \citeauthor{Hahn2015}, which allows them to reduce the fraction of metastable ions in the stored ion beam.

Figure~\ref{fig:Fig10} shows the total cross-section calculated as the sum of 3s shell data of \cite{Younger1983} for \ion{Sc}{X} with the rest of the inner-shell contributions. In this case, the main contributor to the total DI cross-section is the 3s shell followed by the 2p shell in a very low proportion. The EA contribution was added  from \cite{Sampson1982} after applying a scaling factor, as explained in the following section.
As in the case of the Na-like sequence, the total cross-section compared to D07 is systematically10-40\% higher, probably for the same reason.
 
\subsubsection{EA contribution}

 For the Mg-sequence, we  compared the EA contributions calculated with the method explained in Section 2.2 to other calculations available for $Z=$13, 16, 17 and 18 \citep{Tayal1986}, and $Z=$28 \citep{Pindzola1991}.  We have compared Sampson's cross-sections $Q_{\rm SG}$ at about twice the EA onset towards these other calculations $Q_{\rm TP}$, and have found the following relation for these elements. We assume the same relation for $Z>$14 of the Mg-sequence:
\begin{equation}
Q_{\rm TP} = [-0.07 + 0.03306~Z]Q_{\rm SG}.
\end{equation} 
For \ion{Al}{II} and \ion{Si}{III} the scaling factor is smaller:  0.20.  The observations for neutral \ion{Mg}{I} \citep{Freund1990,McCallion1992a}  show no evidence for EA and therefore we neglected this process for neutral Mg.
The available measurements  \citep{Chung1996,Bernhardt2014} show EA enhancements that are consistent with the above scaling.

\subsection{Al isoelectronic sequence}

\subsubsection{DI: 3s cross-sections}

For the Al-sequence up to the Ar-sequence the 3s inner-shell contribution is interpolated from the theoretical calculations of \cite{Younger1982a} for Ar, Sc, and Fe ions. For the P-like and S-like sequences, we  included the data for Ni ions from \cite{Pindzola1991} because they correctly follow  the trend of the rest of the elements in the same sequence, which is not the case for the other sequences. 

\subsubsection{DI: 3p cross-sections}

 For \ion{Fe}{XIV}, the recent measurements of \cite{Hahn2013} using the TSR ion ring storage confirm the existence of a considerably lower cross-section than previous measurements \citep{Gregory1987} or calculations (AR; D07). Hahn's results agree with Gregory's from threshold up to 700~eV, and after that they decrease until they show a difference of 40\%. One of the reason for this difference could be the presence of the metastable ions in Gregory's experiment. The major discrepancies with the theory could come from the fact that theory overestimates the EA component, specially the $n=2\rightarrow4$ transitions, in the case of D07. 

\subsubsection{EA contribution}

For the Al to Ar-sequences, the EA calculations of \cite{Pindzola1991} for Ni ions can be used for comparison with the EA parameters derived from \cite{Sampson1982}.  The scaling factors needed to bring the Sampson data in accordance with the Pindzola data are given in Table~\ref{tab:title}. These data show that the scaling factor gets smaller for higher sequences.  This is no surprise since Sampson's calculations were, in particular, designed for the Na-sequence. We note that the relative importance of the EA process diminishes anyway for the higher sequences. 

Lacking other information we assumed that, for all other ions of these isoelectronic sequences, the same scaling factors apply as for the Ni-ions.  Where there are measurements available with clear indications of the EA process, this scaling appeared to be justified.
The possible exception is \ion{Ni}{XIII} (S-sequence), where \cite{Pindzola1991} suggest that there is an additional contribution in the measurements of \cite{Wang1988} owing to resonant recombination followed by double autoionization. 
However, we  decided to apply the same process as explained above for calculating the scaling factor of the S-sequence, only taking into consideration  the Pindzola data.

\begin{table}[b!]
\caption {EA scaling factors for the Al--Ar isoelectronic sequences needed to bring the Sampson data in accordance with the Pindzola data.}
\label{tab:title} 
\begin{center}
\begin{tabular}{c c}
\hline
\hline
Isoelectronic &Scaling\\
sequence & factor \\[5pt]
\hline
 Al & 0.79 \\
 Si & 0.81 \\
 P  & 0.75 \\
 S  & 0.62 \\
 Cl & 0.64 \\
 Ar & 0.52 \\[5pt]
 \hline
\end{tabular}
\end{center}
\end{table}

\subsection{Si isoelectronic sequence}

\subsubsection{DI: 3p cross-sections}

The experimental data available for \ion{Ar}{V} \citep{Crandall1979,Muller1980,Sataka1989} agree well below 200~eV but, above this energy, the \citeauthor{Crandall1979} data are slightly higher. The three data sets are about 20\% larger than expected based upon Younger's calculations,
probably due to some contamination by metastable levels in the beam. For this reason, theoretical calculations were obtained for \ion{Ar}{V}, taking the $A$, $B$, $C,$ and $D$ parameters proposed by AR for Younger's formula.The same situation occurs for \ion{Cr}{XI} \citep{Sataka1989} and the data were discarded.  

The measurements of Hahn \citep{Hahn2011b,Hahn2012b} are used for \ion{Fe}{XIII}. These data are 20\% lower than the \cite{Arnaud1992} calculations and 15\% lower above $\sim$680~eV, compared with the FAC calculations of D07. The Hahn et al. experimental data show a faster increase of the cross-section in the onset compared with the calculations, probably owing to the excitation of the 3s shell electron, which the calculations did not include. The possible explanation for the higher EA contribution above the threshold proposed by Hahn is that the calculations overestimate the branching ratio of the autoionization and, additionally, the intermediate states could decay by double ionization rather than single ionization.
\begin{figure}
\includegraphics[width=0.55\textwidth]{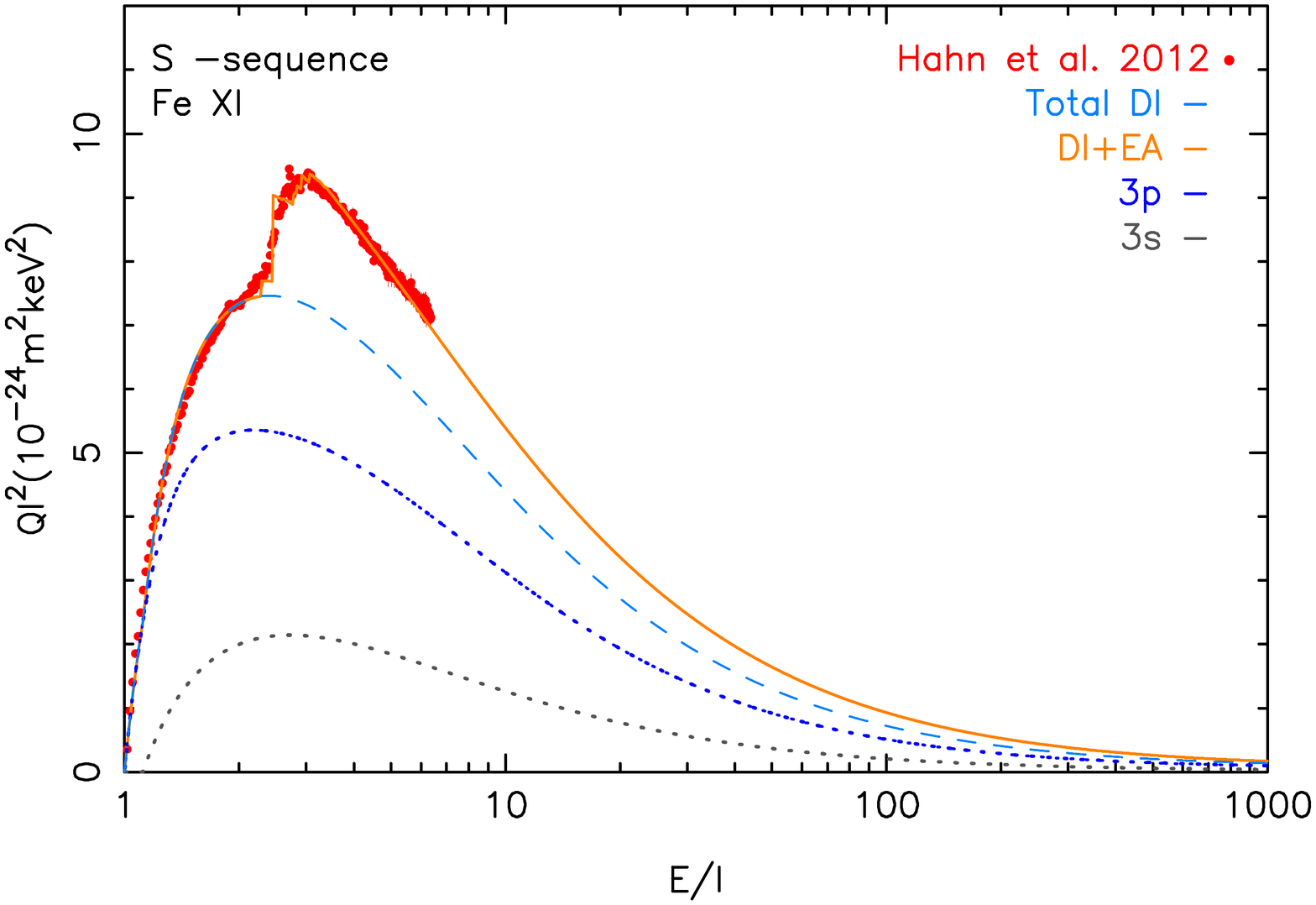}
\caption{Total normalized cross-section for \ion{Fe}{XI} (orange line), DI cross-section (dashed blue line), 3p (dotted blue line), 3s (dotted grey line) and the measurements of \cite{Hahn2012c} (red dots).}
\label{fig:Fig11}
\end{figure}
\subsection{P isoelectronic sequence}

\subsubsection{DI: 3p cross-sections}

The measurements of \cite{Freund1990} for \ion{P}{I} are used for the 3p cross-section fitting with an error less than 6\%. In the case of \ion{S}{II}, two data set are available, \cite{Yamada1988} and \cite{Djuric1993a}, which agree within $\pm15\%$. Yamada's measurements extent up to $u=12$. However, the cross-section at the peak appears about $\sim40-50\%$ larger than expected base on the general trend of the elements in this sequence. The measurements are probably affected by metastable ions in the beam. For this reason, we  decided to use interpolation for this element. 
The measurements of \cite{Hahn2011a} are used for \ion{Fe}{XII} instead of \cite{Gregory1983} because the latter data are compromised owing to metastable ions in the beam. Hahn's data are about $\sim30\%$ lower than the data of \cite{Gregory1983} and the calculations of \cite{Arnaud1992}, and are in agreement with the theoretical cross-section of D07 within $\pm20\%$.

\subsection{S isoelectronic sequence}

\subsubsection{DI: 3p cross-sections}

For \ion{Ar}{III}, we  followed the discussion in \cite{Diserens1988} and did not include the data of \cite{Muller1980}, \cite{Muller1985a}, and \cite{Danjo1984},whose data are larger at high energies than the presently adopted data of \cite{Diserens1988} and \cite{Man1993}. As explained by Diserens, the increased cross-sections may indicate the presence of metastable ions in the beam.

The measurements of \cite{Hahn2012c} for \ion{Fe}{XI} are about 35\% lower than the \cite{Arnaud1992} theoretical calculations and are in reasonable good agreement with D07. The main differences are twofold. First, at 650~eV, a step appears in the cross-section owing to $n=2\rightarrow3$ excitations not included in D07; and secondly, for higher energies D07 considers the $n=2\rightarrow4$ and $n=2\rightarrow5$ EA transitions, resulting in a higher cross-section. However, the experiments do not show evidence for these last processes.

\subsection{Cl isoelectronic sequence}

\subsubsection{DI: 3p cross-sections}

For \ion{K}{III} and \ion{Sc}{V}, the theoretical calculations of \cite{Younger1982c} for the 3p shell were used. For \ion{Ni}{XII}, the measurements of \cite{Cherkani-Hassani2001} and the calculations of \cite{Pindzola1991} seem to be in good agreement, see Fig.~\ref{fig:Fig12}.

The measurements of \cite{Hahn2012c} are used for \ion{Fe}{X}. These are 35\% lower than the \cite{Gregory1987} measurements. The theoretical calculations of \cite{Arnaud1992} and D07 lie within the experimental uncertainties, although some discrepancies can be found owing to the non-identical EA processes modeling. The reason for these differences are the same as for \ion{Fe}{XI}, explained in section 3.16.1.
\begin{figure}
\includegraphics[width=0.55\textwidth]{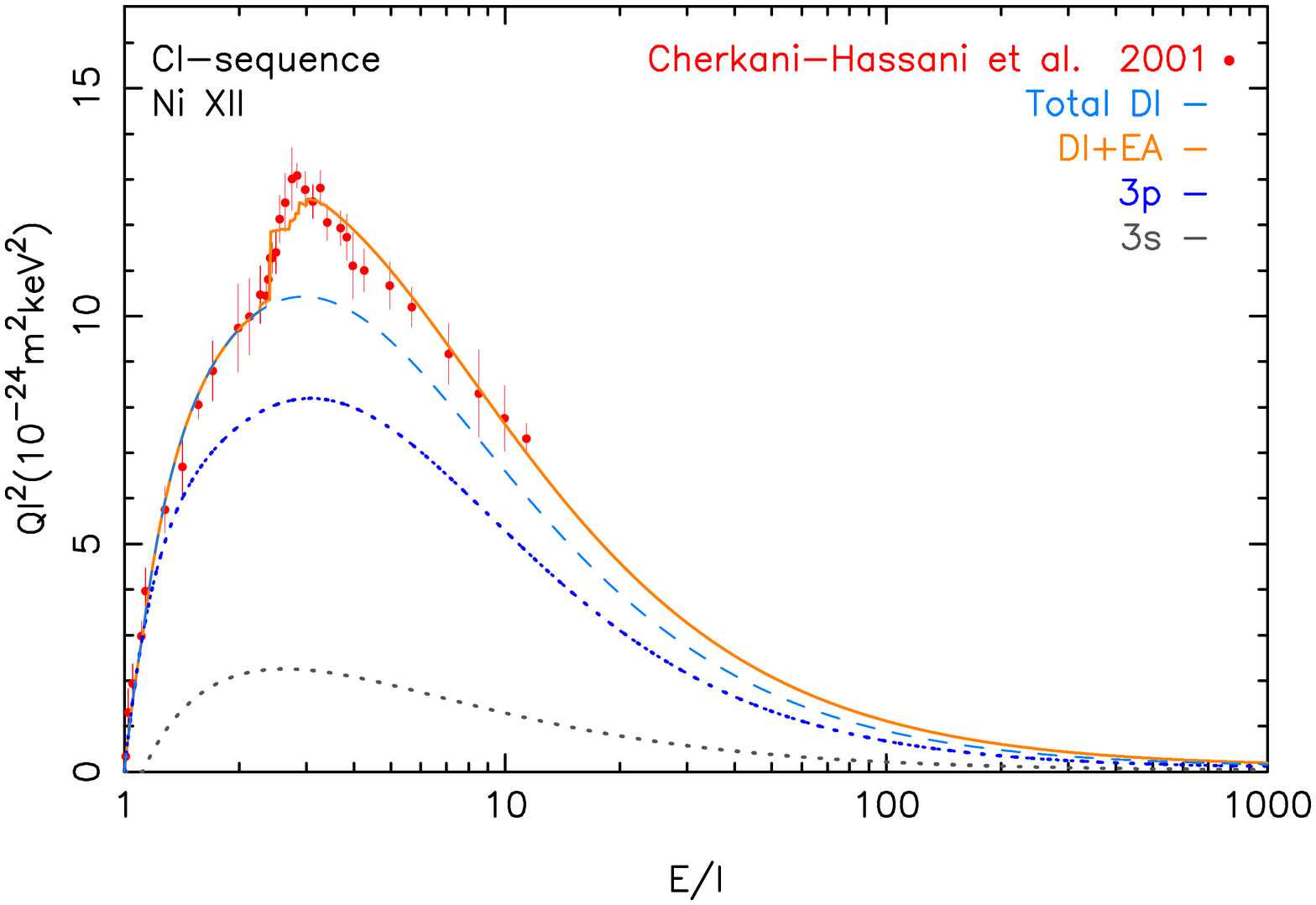}
\caption{Total normalized cross-section for \ion{Ni}{XII} (orange line), DI cross-section (dashed blue line), 3p (dotted blue line), 3s (dotted grey line) and the measurements of \cite{Cherkani-Hassani2001} (red dots).}\label{fig:Fig12}
\end{figure}

\subsection{Ar isoelectronic sequence}
\subsubsection{DI: 3p cross-sections}

The theoretical data of \cite{Younger1982d} for the 3p shell of \ion{Sc}{IV} were  taken into account, which are in good agreement with the D07 FAC calculations. Otherwise, for \ion{Fe}{IX} recent measurements of \cite{Hahn2016} are available. In this case, the storage ring could not eliminate all the metastable ions from the beam. However, \citeauthor{Hahn2016} are able to estimate a metastable fraction of 30\% in the 3p$^5$3d level and they obtain a new estimated ground state cross-section (subtracting the metastable states from the experimental data), which is 15-40\% lower than the AR and D07 calculations,  and 20\% lower than the total cross-section derived from the \cite{Younger1982d} data for the 3p shell. Owing to those lower values of the total cross-section, the rest of the elements interpolated or extrapolated based on \ion{Fe}{IX} will be affected as well by a systematic decrease of their total cross-section.

\subsection{K isoelectronic sequence}

The K-like (3s$^2$3p$^6$4s) ions have the 3p and 3d shells as the main contributors to the DI and the EA process is dominated by excitation from 3p$^6$3d to the 3p$^5$3d$nl$ levels with $n=4,5$. The DI contribution of 4s is taken into account for the elements that have some electrons in the 4s shell, such as \ion{K}{I} and \ion{Ca}{II} with a structure of 3s$^2$3p$^6$4s. The same process has been followed for the ions up to the Zn-like sequence that have the 4s shell contribution. 

\subsubsection{DI: 3s \& 3p cross-sections}

For the calculation of the 3s and 3p shells DI cross-section contribution, we have followed the same procedure as for the 2s and 2p shells, explained in section 3.12.1. We  calculated the $A$, $B$, $D,$ and $E$ parameters with Eq. \ref{eqn:interpar} using, as reference, the parameters of the previous isoelectronic sequence. The same process was applied for all elements from the K-sequence up to the  Zn-sequence.

\subsubsection{DI: 3d \& 4s cross-sections}

For \ion{K}{I} and \ion{Ni}{X} we  used the theoretical data of \cite{McCarthy1983} and \cite{Pindzola1991}, respectively, which are well fitted. For \ion{Fe}{VIII,}  the recent measurements of \cite{Hahn2015} were used, from the ionization threshold up to 1200~eV. They remain 30-40\% lower than theoretical calculations of \cite{Arnaud1992}, based on \cite{Pindzola1987}, and are in good agreement (10\%) with D07. The reason for these discrepancies are similar to the case of \ion{Fe}{XI}, as explained in section 3.16.1.

\subsubsection{EA contribution}
We  adopted the EA parameters calculated by D07 from his FAC EA calculations, which are the same as used by CHIANTI for all the sequences from the K-like up to the Cr-like sequences.

\subsection{Ca isoelectronic sequence}

\subsubsection{DI: 3d \& 4s cross-sections}

For \ion{Ca}{I} we  selected three data sets \citep{McGuire1977,McGuire1997,Roy1983} of theoretical calculations. The first two sets of McGuire are in reasonably good agreement, although they are $\sim$30\% higher than Roy's. There are no apparent reasons for discarding any of the three sets and therefore we  decided to include all of them. For \ion{Fe}{VII} we use the sets of \cite{Gregory1986} and \cite{Stenke1999}  and for\ion{Ni}{IX} the calculations of \cite{Pindzola1991} and \cite{Wang1988}.

\subsection{Sc isoelectronic sequence}

\subsubsection{DI: 3d \& 4s cross-sections}

We include measurements of \ion{Ti}{II}, \ion{Fe}{VI,} and \ion{Ni}{VIII} to obtain the DI and EA cross-sections of the scandium (3p$^6$3d$^3$) sequence. The rest of the elements are interpolated or extrapolated. For \ion{Fe}{VI}, the measurements of \cite{Gregory1987} and \cite{Stenke1999} are in good agreement with our fit. The data sets of \cite{Wang1988} and \cite{Pindzola1991} were  used for \ion{Ni}{VIII}.

\begin{figure}
\includegraphics[width=0.55\textwidth]{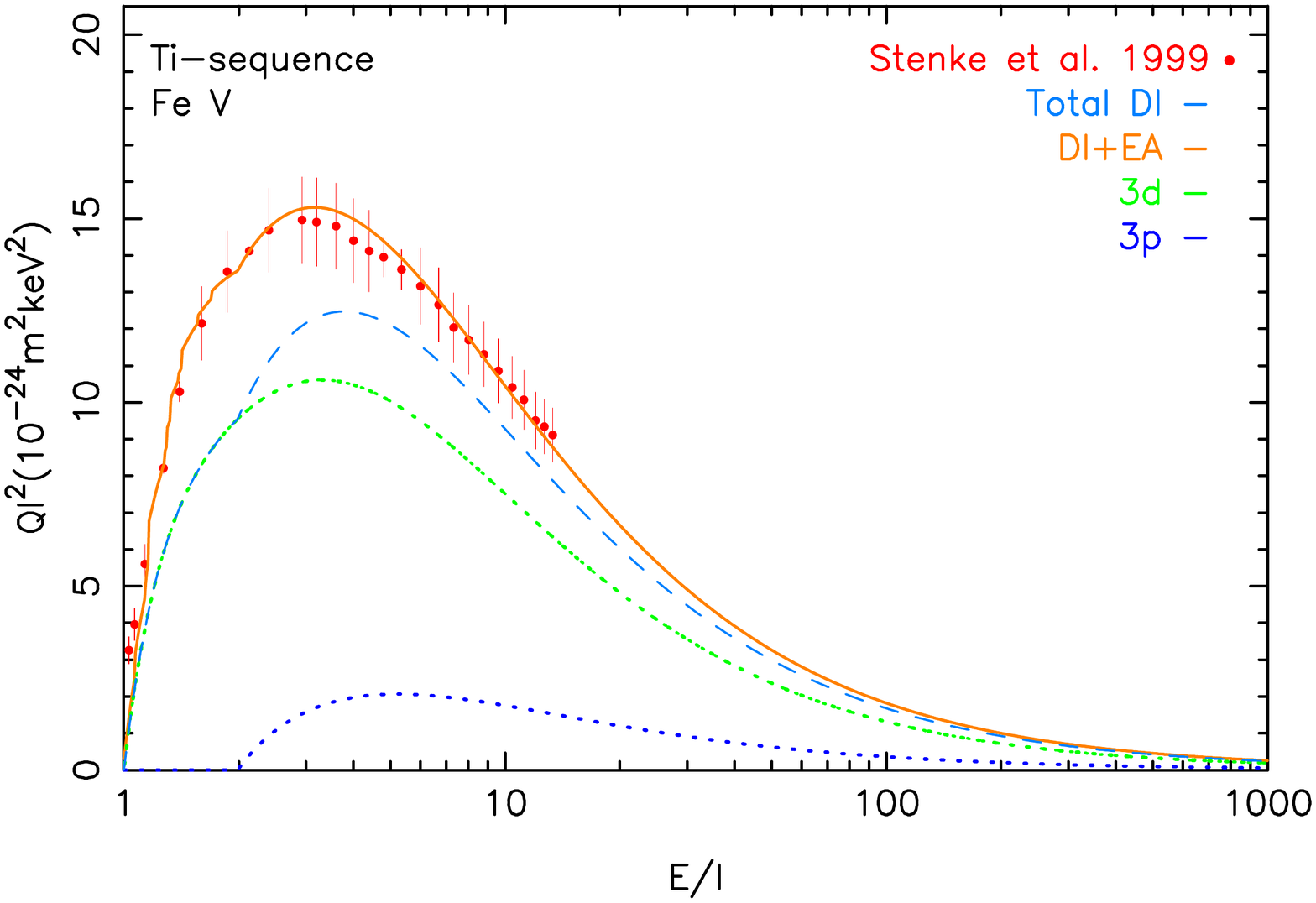}
\caption{Total normalized cross-section for \ion{Fe}{V} (orange line), DI cross-section (dashed blue line), 3d (dotted blue line), 3p (dotted blue line) and the measurements of \cite{Stenke1999} (red dots).}
\label{fig:Fig12}
\end{figure}

\subsection{Ti isoelectronic sequence}

\subsubsection{DI: 3d \& 4s cross-sections}
For the titanium sequence (3p$^6$3d$^4$) the measurements of \cite{Stenke1999} for the \ion{Fe}{V}, as can be seen in Fig.~\ref{fig:Fig12}, and \ion{Ni}{VII} 3d shell, respectively, were  selected. In the case of \ion{Ti}{I}, with an irregular structure of 3d$^2$4s$^2$, the \cite{McGuire1977} theoretical calculations are used to obtain the 4s shell contribution.

\subsection{V isoelectronic sequence}

\subsubsection{DI: 3d \& 4s cross-sections}

The measurements used for the V-like sequence are \cite{Tawara2002} for \ion{V}{I} (obtained directly from CHIANTI), \cite{Man1987a} for \ion{Cr}{II}, \cite{Stenke1999} for \ion{Fe}{IV} and \cite{Wang1988} for \ion{Ni}{VI}, which are well fitted. The remaining elements are interpolated or extrapolated.

\subsection{Cr isoelectronic sequence}

\subsubsection{DI: 3d \& 4s cross-sections}

For \ion{Cr}{I} there are no measurements available and we use the calculations of \cite{Reid1992} and \cite{McGuire1977} for high energies.  The measurements of \cite{Bannister1993} and calculations of \cite{Pindzola1991} are in good agreement for \ion{Ni}{V}.

\subsection{Mn isoelectronic sequence}

\subsubsection{DI: 3d \& 4s cross-sections}

For the manganese sequence (3p$^6$3d$^7$) elements 3d shell DI cross-section, we use the theoretical calculation of \cite{Younger1983} for \ion{Fe}{II} and the measurements of \cite{Gregory1986} for \ion{Ni}{IV}. Since the \ion{Fe}{II} element has a ground state configuration of 3d$^6$4s, we  considered the measurements of the total DI cross-section of \cite{Montague1984a} for subtracting the contribution of the rest of the inner-shells and for obtaining the 4s shell DI cross-sections. \ion{Mn}{I} was  fitted with data of \cite{Tawara2002} taken from CHIANTI (D07). The remaining elements of the sequence are interpolated. 

\subsection{Fe isoelectronic sequence}

\subsubsection{DI: 3d \& 4s cross-sections}
 
 For \ion{Fe}{I,} we  included the measurements of \cite{Freund1990} and the FAC DI calculations of D07 for \ion{Co}{II}. We use the \cite{Pindzola1991} theoretical calculations for \ion{Ni}{III}, which are in good agreement with \cite{Stenke1999} at high energies; and \cite{Gregory1986} for \ion{Cu}{IV}. The rest of the elements of the sequence are interpolated. 

\subsection{Co isoelectronic sequence}

\subsubsection{DI: 3d \& 4s cross-sections}

The measurements found for the cobalt sequence are \cite{Montague1984a} for \ion{Ni}{II} and \cite{Gregory1986} for \ion{Cu}{III}, which are well fitted. \ion{Co}{I} was  fitted with data of \cite{Tawara2002} taken from CHIANTI (D07)  and \ion{Zn}{IV} with the extrapolation of \ion{Ni}{II} and \ion{Cu}{III}.

\subsection{Ni isoelectronic sequence}
\subsubsection{DI: 3d \& 4s cross-sections}
For \ion{Ni}{I}  (with ground configuration 3d$^8$4s$^2$) the \cite{Pindzola1991} and \cite{McGuire1977} data were selected for the 3d and 4s DI contribution, respectively. For \ion{Cu}{II} and \ion{Zn}{III}, there are no known measurements, therefore, the 3d DI cross-sections were  calculated as the extrapolation of Pindzola's data for \ion{Ni}{I}.

\subsection{Cu isoelectronic sequence}

\subsubsection{DI: 3d \& 4s cross-sections}

 The measurements considered for the 4s shell DI fit of the copper sequence (3d$^{10}$4s) are for \ion{Cu}{I}, \cite{Bolorizadeh1994} and \cite{Bartlett2002}; and for \ion{Zn}{II}, \cite{Peart1991a} and \cite{Rogers1982}. The fit is in a reasonably good agreement with the measurements. The 3d shell DI contribution of both elements were  calculated with FAC.

\subsection{Zn isoelectronic sequence}

\subsubsection{DI: 3d \& 4s cross-sections}

For the \ion{Zn}{I} ion, which has a 3d$^{10}$4s$^2$ ground configuration, the calculations of \cite{Omidvar1977} were used for the fit of the 4s DI cross-section contribution instead of \cite{McGuire1977}, because they are around 5-10\% higher than Omidvar's values before the cross-section peak and more than 20\% lower after, which disagrees with the contribution of the inner-shells. Otherwise, FAC was used forthe 3d DI cross-section calculation.

\section{Ionization rate coefficients}

In the previous section, we  obtained the ionization cross-sections for all subshells of all elements from H to Zn isoelectronic sequence by applying Eq. \ref{eqn:younger_ext} for DI and Eq. \ref{eqn:eamewe} for EA. The total cross-section for DI can be written as the sum of $j$ inner-shells, where  $u_j=E_{\rm e}/I_j$ with $E_{\rm e}$ the incoming electron energy (in keV) and
$I_j$ the ionization potential of the atomic subshell (in keV): 
\begin{multline}
Q_{DI} =\sum_j{1\over{u_jI_j^2}} \bigg[A_j\left(1-{1\over u_j}\right) + B_j \left(1 - {1\over u_j}\right)^2 \\+ C_j R_j \ln u_j  + D_j {\ln u_j\over \sqrt{u_j}} +  E_j{\ln u_j\over u_j}\bigg].
\end{multline}

The parameters $A_j$, $B_j$, $C_j$, $D_j$, and $E_j$ (in units of 10$^{-24}$m$^2$keV$^2$) for Si-like \ion{Fe}{XI} are given in  Table E.1.

The direct ionization rate is written as a function of the temperature $kT$ as
\begin{equation}
\displaystyle C_{DI} = {{2\sqrt{2}n_en_i}\over{[\pi(kT)^3m_e]^{1\over2}}}\sum_{i} C_i \cdot g_i(u_j),\\ 
\label{eqn:di_ir}
\end{equation}

where $n_e$ and $n_i$ are the electron and ion density, respectively,  $m_e$ the electron mass, and  $C_i$ and $g_i(u_j)$ are given in Appendix B.
The same approach can be taken with the EA process. The EA cross-section contribution to the outer shell of each ion, is the sum of $k$ energy level transitions with $I_{EAk}$ the excitation-autoionization threshold (in keV):
\begin{equation}
Q_{EA} =\sum_k{1\over{u_kI_{EAk}^2}} \bigg[A_{EAk}+ {B_{EAk}\over u_k}  + {C_{EAk} \over u_k^2} + {2D_{EAk} \over u_k^3} + E_{EAk} \ln u_k\bigg], \
\end{equation}
where  $A_{EAk}$, $B_{EAk}$, $C_{EAk}$, $D_{EAk}$, and $E_{EAk}$ (in units of 10$^{-24}$m$^2$keV$^2$) are the parameters obtained for each ion in presence of the EA process.

Moreover, the total excitation-autoionization rate coefficient is expressed as
\begin{equation}
\displaystyle C_{EA} = {{2\sqrt{2}n_en_i}\over{[\pi(kT)^3m_e]^{1\over2}}}\sum_{i} D_i \cdot m_i(u_k).\\ 
\label{eqn:ea_ir}
\end{equation}

A detailed description of the $D_i$ and $m_i(u_k)$ terms of this parametric formula is shown in Appendix C.

The total ionisation rate coefficient is given by the sum of Eq.\ref{eqn:di_ir} and \ref{eqn:ea_ir} and includes the contributions from all inner-shells.

\begin{figure}
\includegraphics[width=0.50\textwidth]{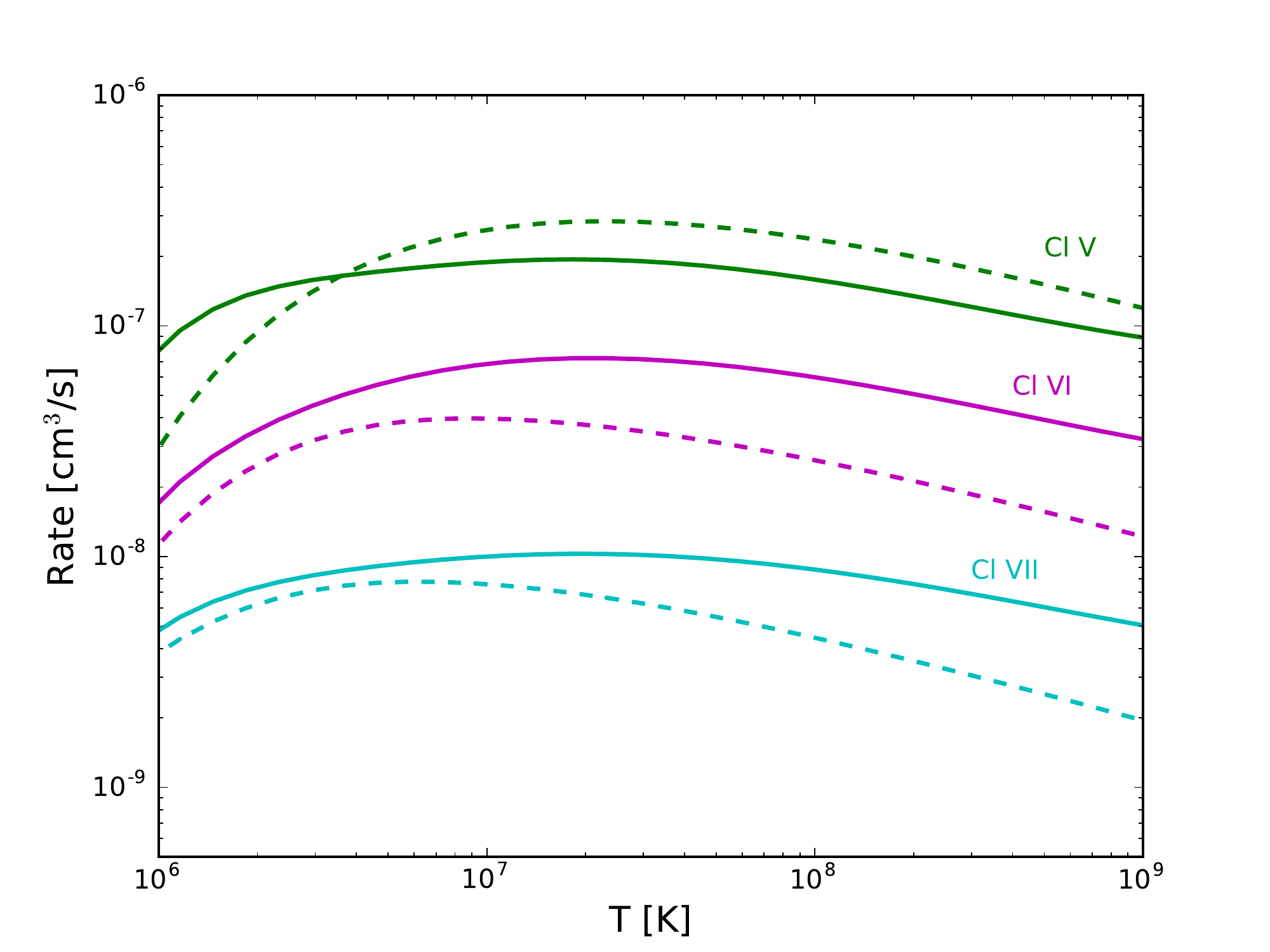}
\caption{Left axis: Ionization rates  comparison between \cite{Bryans2009} (dashed lines) and the present work (solid lines) for \ion{Cl}{VII} (Na-like), \ion{Cl}{VI} (Mg-like, rate multiplied by factor 10) and \ion{Cl}{V} (Al-like, rate multiplied by factor 50).}
\label{fig:Fig13}
\end{figure}

 \begin{figure}
\includegraphics[width=0.50\textwidth]{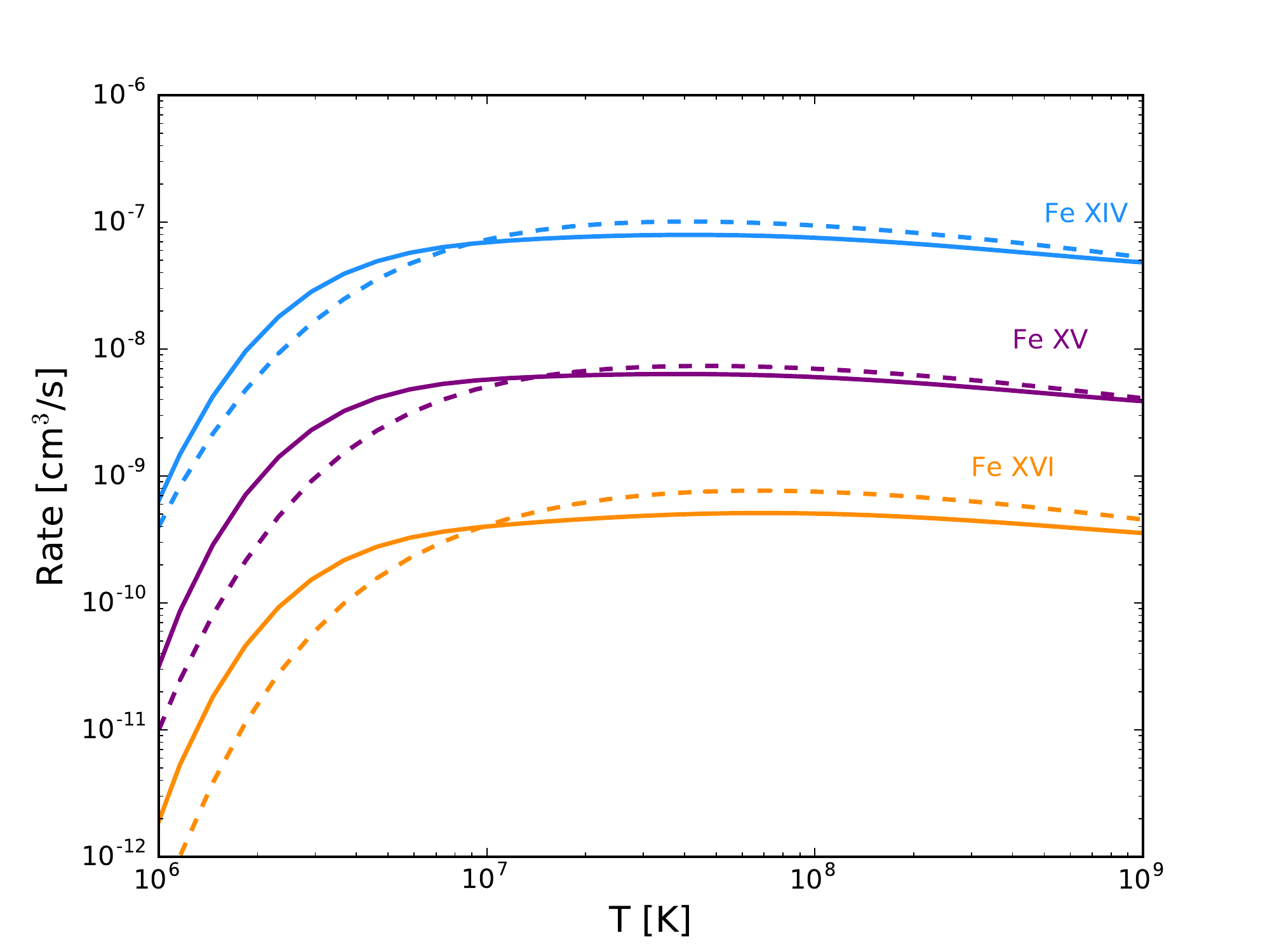}
\caption{Ionization rates comparison between \cite{Bryans2009} (dashed lines) and the present work (solid lines) for \ion{Fe}{XVI} (Na-like), \ion{Fe}{XV} (Mg-like, rate multiplied by factor 10), and \ion{Fe}{XIV} (Al-like, rate multiplied by factor 100).}
\label{fig:Fig14}
\end{figure}

\section{Discussion}

A systematic comparison was made with the \cite{Bryans2009} atomic data, which adopt the D07 electron-impact ionization rates. This shows that the present work and \cite{Bryans2009} rates are in good agreement (differences less than 10-20\%) for more than 85\% of the elements. The highest differences appear for the isoelectronic sequences of Na, Mg (\ion{Si}{III}, \ion{P}{IV}, \ion{S}{V} and \ion{Cl}{VI}), and Al (\ion{P}{III}, \ion{S}{IV}, \ion{Cl}{V} and \ion{Ar}{VI}), where some of the ions 
show a difference of 30-40\% in the cross-sections compared with D07. As a consequence, the ionization rates for these ions are up to 2-3 times higher than D07 for high temperatures. An example for Cl is shown in Fig.~\ref{fig:Fig13}. This difference decreases for high $Z$ elements as can be seen in Fig.~\ref{fig:Fig14} where the ionization rates for Fe are represented. 

A possible explanation could be that the D07 cross-sections are mainly calculated with FAC, instead of fitted to experimental data, as performed in the present work (see Appendix A). The measurements represent a more realistic scenario and include more transitions, since REDA or multiple ionization  are not usually incorporated in the theoretical calculations.

 \begin{figure}
\includegraphics[width=0.50\textwidth]{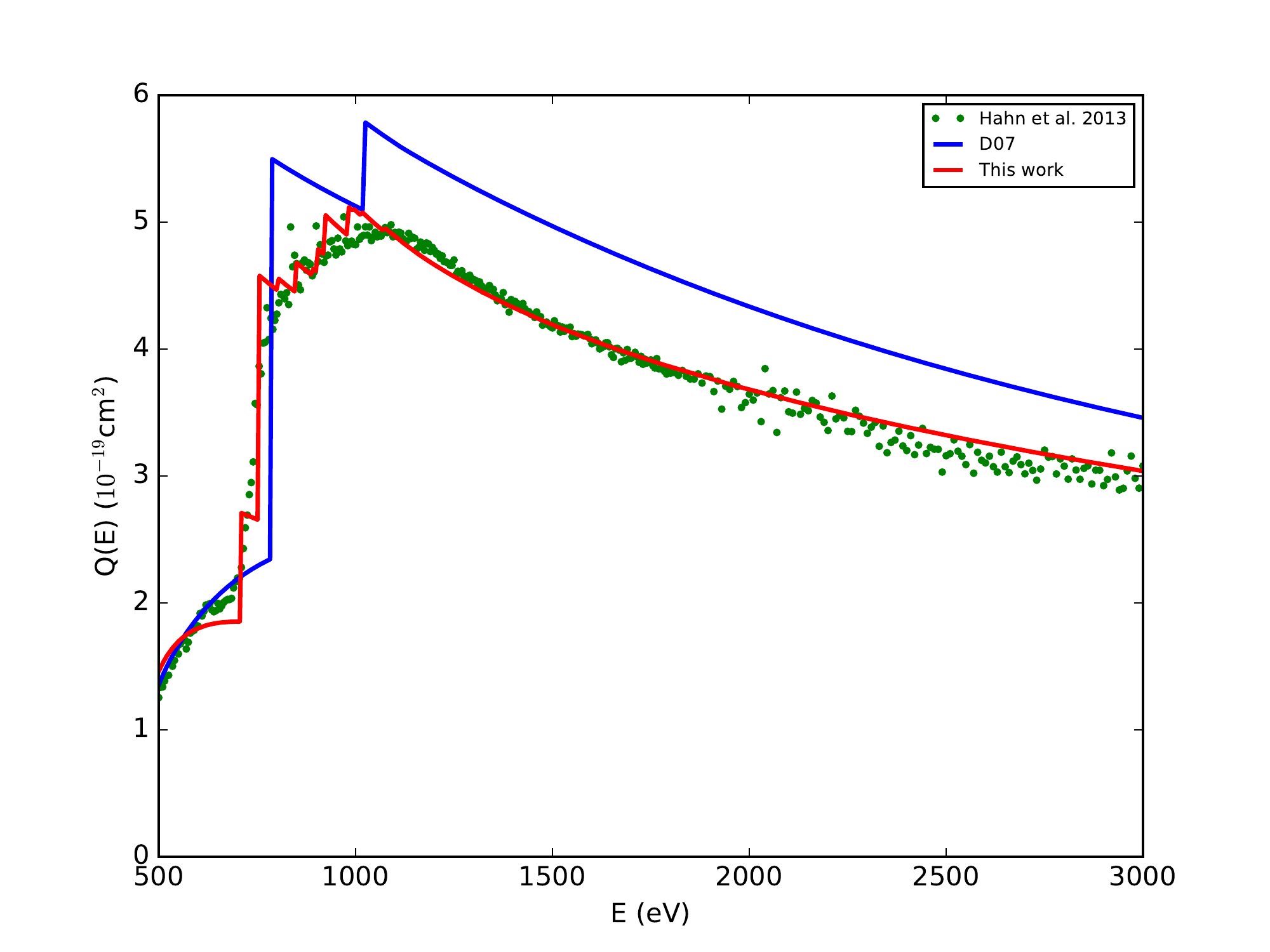}
\caption{\ion{Fe}{XIV} total cross-section. The experimental results of \citeauthor{Hahn2013} are shown by the green dots. The theoretical calculations of D07 are given by the blue line and the results of this work derived from the fitting process described in section 3.13 by the red line. }\label{fig:Fig15}
\end{figure}

 \begin{figure}[h!]
\includegraphics[width=0.50\textwidth]{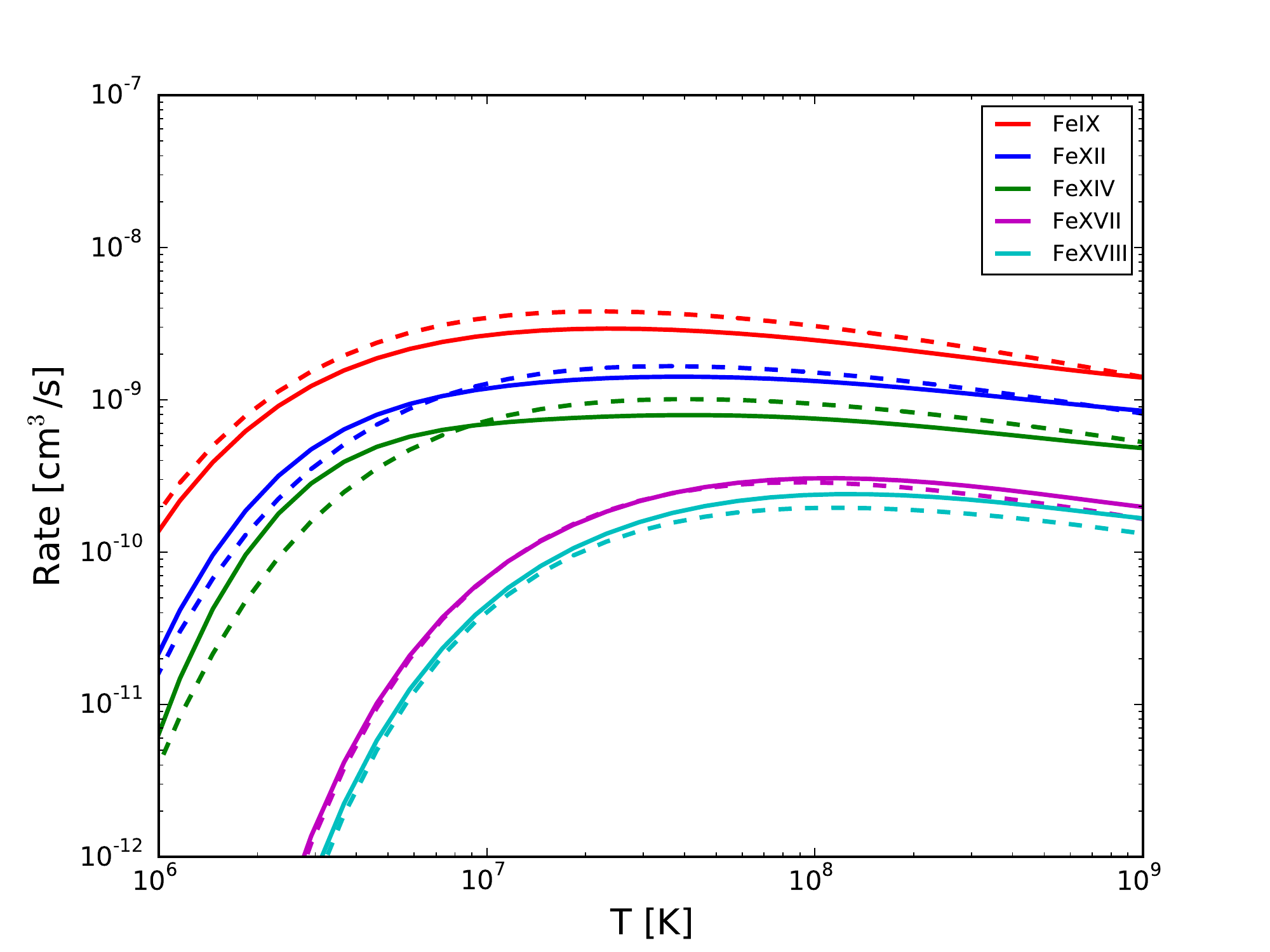}\caption{Ionization rates comparison between \cite{Bryans2009} (dashed lines) and the present work (solid lines).}
\label{fig:Fig16}
\end{figure}

As explained in the previous sections, the most recent experimental measurements included in this work are \ion{Fe}{XVIII}, \ion{Fe}{XVII,} and \ion{Fe}{XIV} \citep{Hahn2013}, \ion{Fe}{XIII} \citep{Hahn2012b}, \ion{Fe}{XII} \citep{Hahn2011a}, \ion{Fe}{XI} and \ion{Fe}{X} \citep{Hahn2012c}, \ion{Fe}{IX} \citep{Hahn2016} and \ion{Fe}{VIII} \citep{Hahn2015}, and \cite{Bernhardt2014} for \ion{Fe}{XV}. They  used the new TSR technique to reduce the metastable ion levels to obtain lower cross-sections than AR for all the ions.
 D07's cross-sections are about 20\% higher than Hahn’s for  \ion{Fe}{XIV}, \ion{Fe}{XIII}, \ion{Fe}{XII}, \ion{Fe}{XI}, \ion{Fe}{X}, \ion{Fe}{IX,} and \ion{Fe}{VIII}. For the other ions, the cross-sections are comparable, with the difference that the  D07 EA threshold is located at higher energy, probably because D07 does not include the $n=2\rightarrow3$ excitations, see Fig.~\ref{fig:Fig15}.
 
 Figure~\ref{fig:Fig16} contains the ionization comparison rates  for some ions: \ion{Fe}{XVIII}, \ion{Fe}{XVII}, \ion{Fe}{XIV}, \ion{Fe}{XII,} and  \ion{Fe}{IX}. The plot shows that the ionization rates are similar or lower than D07, as expected from the experimental measurements. For \ion{Fe}{XI} to \ion{Fe}{XV} we obtain a higher value than D07 for low temperature. The reason for this is probably that, at low temperature, the rates are very sensitive to the weighting of the cross-section with the Maxwellians velocity and a small variance in the cross-section fit at low energies could have a major impact in this region.

The major impact of the new cross-sections used in this work is on the ion fractions obtained by the ionization balance. As an example, we  compared the ion fractions from  \cite{Bryans2009} with the present work for all ions of Fe. In this comparison, we  used the same recombination rates as \cite{Bryans2009}, so the only differences are the ionization rates.

\begin{figure}[h!]
\centering
\begin{minipage}[t]{1\textwidth}
\includegraphics[width=0.50\textwidth]{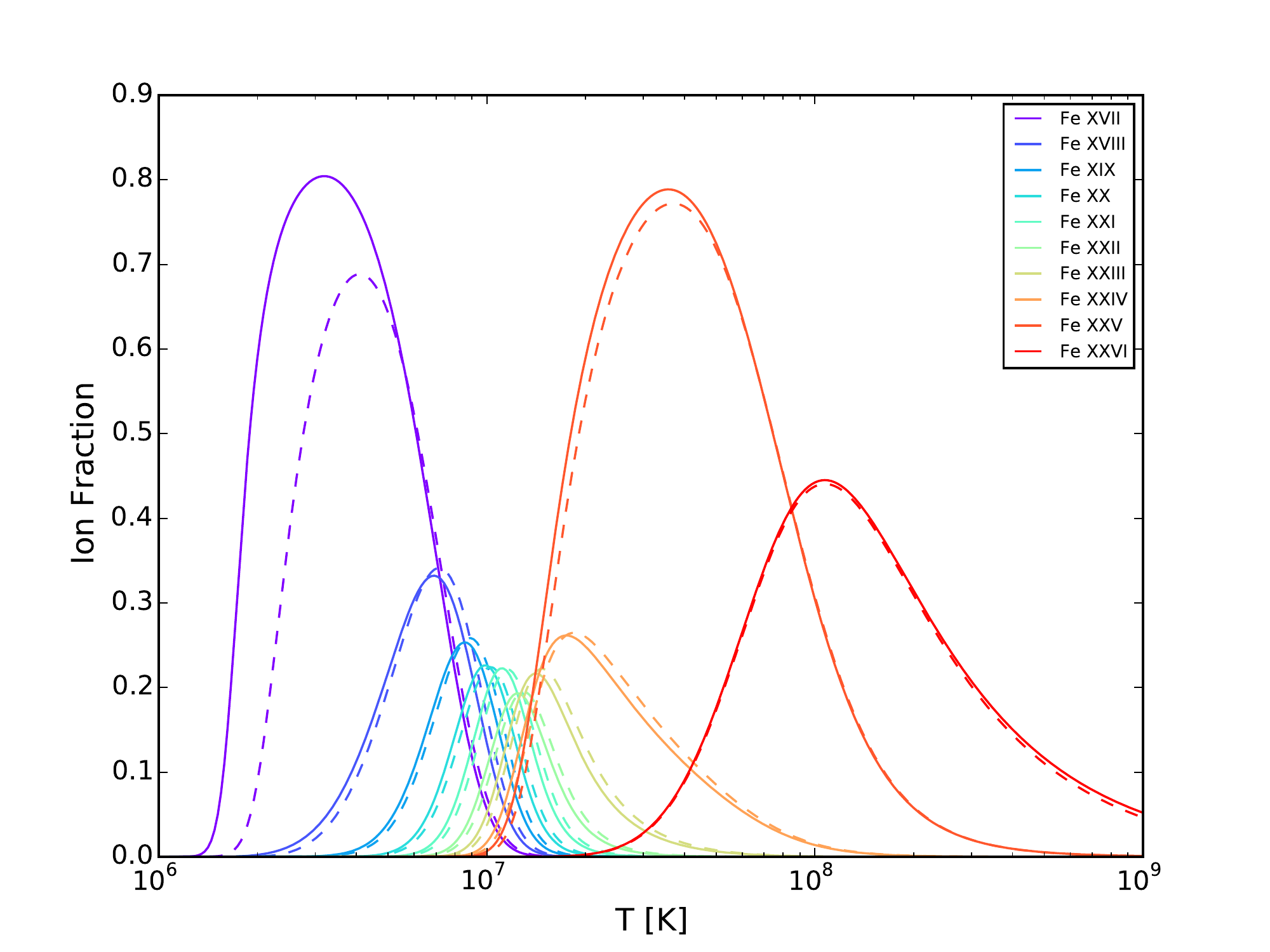}
\end{minipage}
\hfill
\begin{minipage}[t]{1\textwidth}
\includegraphics[width=0.50\textwidth]{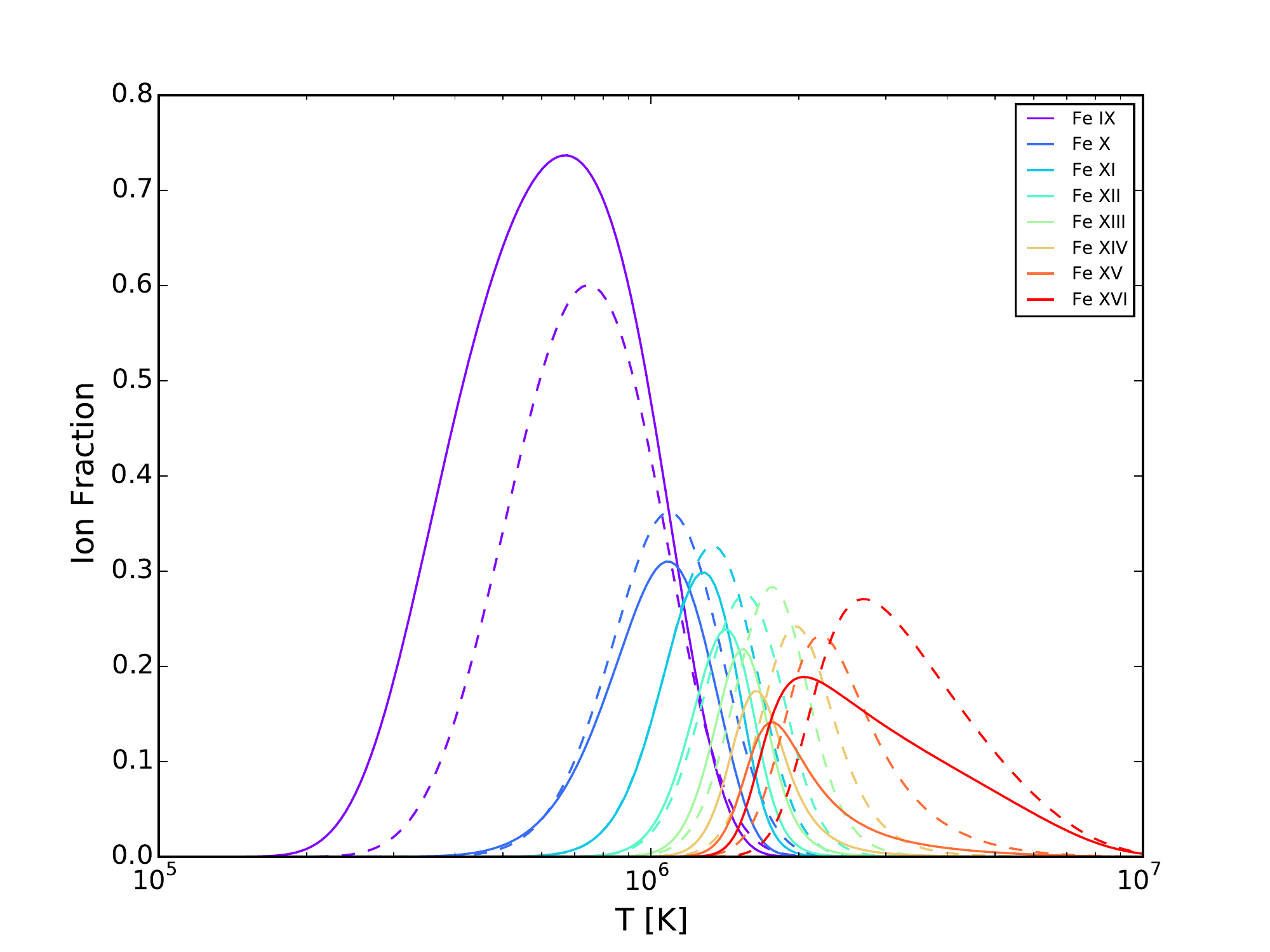}
\end{minipage}
\hfill
\begin{minipage}[t]{1\textwidth}
\includegraphics[width=0.50\textwidth]{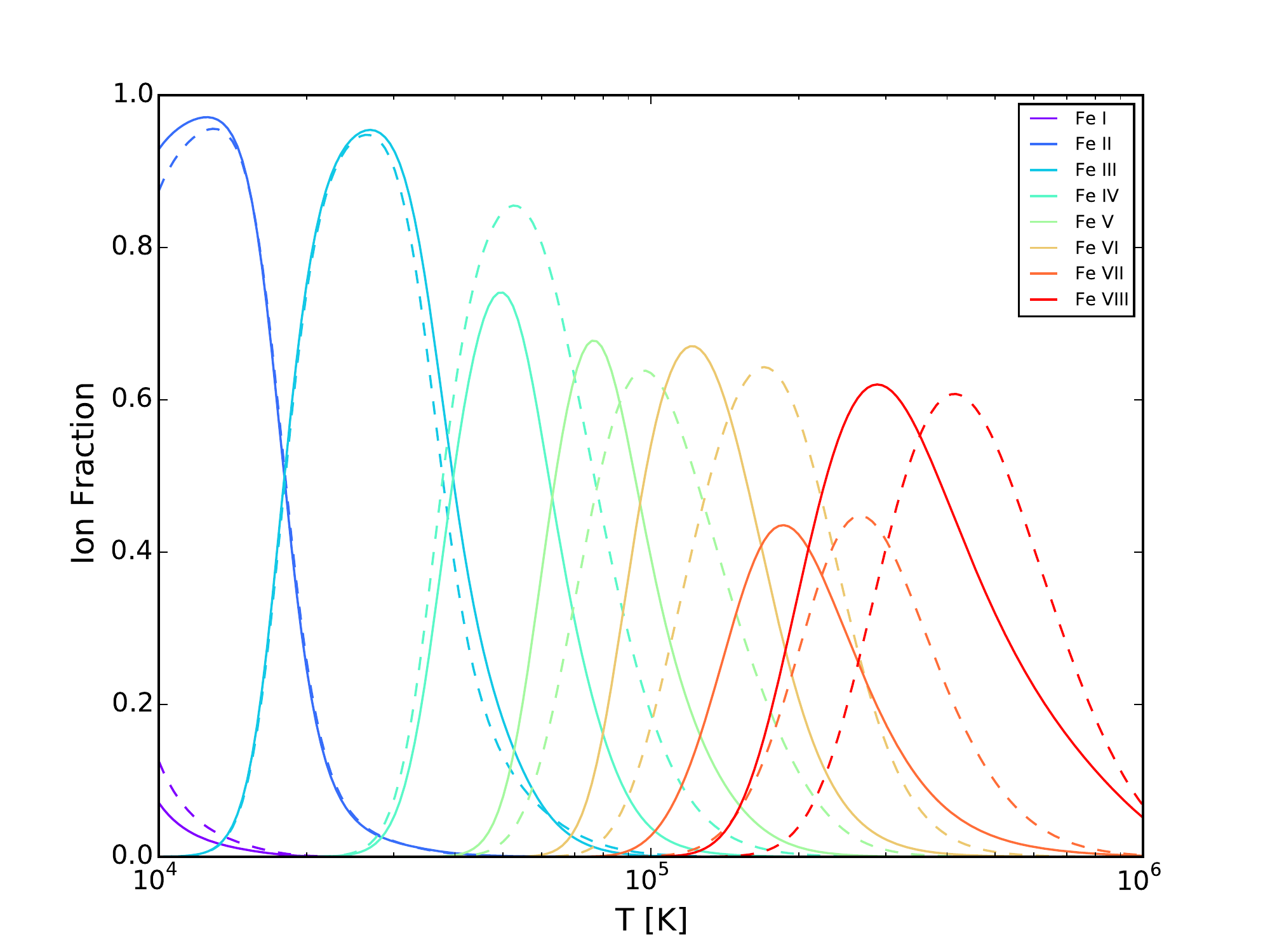}
\end{minipage}
\caption{Top: Ion fraction for H-like to Ne-like Fe including the ionization rates of \cite{Bryans2009} (dashed lines) and present work (solid lines). Middle: Ion fraction for Na-like to Ar-like. Bottom: Ion fraction for K-like to Fe-like.}
\setlength{\unitlength}{5cm}
\label{fig:Fig17}
\end{figure} 

Fig.~\ref{fig:Fig17} (top) shows the first ten ions from \ion{Fe}{XXVI} to \ion{Fe}{XVII}. The ion fraction is relatively similar for all ions, except for Ne-like \ion{Fe}{XVII}. The lower temperature ionic fraction in this work is clearly higher than using \cite{Bryans2009}. This is mainly influenced by ions of adjacent isoelectronic sequences such as Na-like or Mg-like, which have higher ionization rates, as explained above.

Figure~\ref{fig:Fig17} (middle) presents the ion fraction for \ion{Fe}{IX} up to \ion{Fe}{XVI}. In this case, there is a more appreciable difference. The peak ion concentration in the present work is lower than in \citeauthor{Bryans2009} and it seems to be slightly displaced to lower temperatures. However, for \ion{Fe}{XVI} the ion concentration behaviour is similar to \ion{Fe}{XVII}. The least ionized Fe (\ion{Fe}{VIII} up to \ion{Fe}{I}) ion fractions are plotted in Fig.~\ref{fig:Fig17} (bottom). From \ion{Fe}{VIII} to \ion{Fe}{VI,} the values at the peak of the ion fractions are similar but they are displaced at lower temperatures around $\sim$10$^4$-10$^5$ K. For \ion{Fe}{III,} the value at the peak is $\sim$20\% lower. The \ion{Fe}{I} and \ion{Fe}{II} ions are in good agreement.

\section{Summary and conclusions}

We produced a complete set of electron direct collisional ionization cross-sections together with excitation-autoionization cross-sections. We were able to obtain not only the total cross-sections, such as D07, but 
all the individual inner shells cross-section of all elements from the H to Zn isoelectronic sequences. They were obtained from experimental measurements, theoretical calculations, and interpolation/extrapolation among the data sets. We  incorporated the most recent experimental measurements available at the moment, taken by \cite{Fogle2008}, \cite{Hahn2011a,Hahn2011b,Hahn2012a,Hahn2012b,Hahn2012c, Hahn2013,Hahn2015,Hahn2016}, and  \cite{Bernhardt2014}. 

This method enables a much more efficient analytical calculation of ionization rate coefficients than other plasma codes with a comparable accuracy.  The corresponding rates are in good agreement with \cite{Bryans2009} in at least 85\% of the cases. This capability is essential to resolve emission lines and line fluxes in a high-resolution X-ray spectra.

 The results of the present work are included in the  SPEX\footnote{https://www.sron.nl/spex} \citep{Kaastra1996} software, utilized for X-ray spectra modeling, fitting, and analysis. 

\begin{acknowledgements}
SRON is supported financially by NWO, the Netherlands Organization for Scientific Research. We thank J. Mao, L. Gu, T. Raassen and J. de Plaa for their support in the different stages of this project. 
\end{acknowledgements}



\newpage
\tiny
\begin{appendix}
\section{References for used cross-section data}

\begin{threeparttable}
\begin{tabular}{l c p{4cm} c}

\hline
\hline
Ion & Type \tnote{1} & Reference & Uncertainty\\[5pt]
\hline
& &{\bfseries H-sequence} \\
\hline
\ion{H}{I}      & e & \cite{Shah1987} & 7\%              \\
\ion{He}{II}    & e & \cite{Peart1969} & 12\%          \\
\ion{Li}{III}   & e & \cite{Tinschert1989}& 10\%        \\
\ion{B}{V}      & e & \cite{Aichele1998}& 10\%          \\
\ion{C}{VI}     & e & \cite{Aichele1998}& 10\%          \\
\ion{N}{VII}    & e & \cite{Aichele1998} & 10\%         \\
\ion{O}{VIII}   & e & \cite{Aichele1998}& 10\%          \\
\ion{Ne}{X}     & t & \cite{Fontes1999} & --                \\
                & t & \cite{Kao1992} & --                 \\
\ion{Fe}{XXVI}  & t & \cite{Kao1992}& --\\
         & t & \cite{Moores1990}& --       \\
\ion{Cu}{XXIX}  & t & \cite{Moores1990} & --         \\[5pt]
 
\hline
&& {\bfseries He-sequence } \\
\hline
\ion{He}{I}     & e & \cite{Montague1984b}&4\%               \\
         & e & \cite{Rejoub2002} & 5\%                  \\
         & e & \cite{Shah1988} & 6\%                  \\
\ion{Li}{II}    & e & \cite{Peart1968} & 6\%           \\
         & e & \cite{Peart1969} &11\%                 \\
\ion{B}{IV}     & e & \cite{Crandall1979}&4\%               \\
\ion{C}{V}      & e & \cite{Crandall1979} &7\%              \\
         & e & \cite{Donets1981}&6\%          \\
\ion{O}{VII}    & t & \cite{Zhang1990}&--               \\
\ion{Ne}{IX}    & e & \cite{Duponchelle1997}&11\%             \\
         & t & \cite{Zhang1990}&--               \\
\ion{Fe}{XXV}   & t & \cite{Zhang1990}&--              \\
\ion{Zn}{XXIX}  & t & \cite{Zhang1990}&--              \\[5pt]
 
\hline
& & {\bfseries Li-sequence (1s)} \\
\hline
\ion{Li}{I}     & t & \cite{Younger1981a}&--                      \\
\ion{Be}{II}    & t & \cite{Younger1981a}&--                      \\
\ion{O}{VI}     & t & \cite{Zhang1990}&--              \\
\ion{Fe}{XXIV}  & t & \cite{Zhang1990}&--             \\
\ion{Zn}{XXVIII}  & t & \cite{Zhang1990}&--             \\[5pt]
 
\hline
& &  {\bfseries Li-sequence (2s)} \\
\hline
\ion{Li}{I}     & t & \cite{Bray1995} &--                  \\
         & e & \cite{Jalin1973}& 18\%                \\
\ion{Be}{II}    & e & \cite{Falk1983b}& 8\%                 \\
\ion{B}{III}    & e & \cite{Crandall1986}& 7\%             \\
\ion{C}{IV}     & e & \cite{Crandall1979}& 7\%                   \\
                & e & \cite{Teng2000}& 14\%                   \\
\ion{N}{V}      & e & \cite{Crandall1979}& 7\%                  \\
           & e & \cite{Donets1981}&--          \\
\ion{O}{VI}     & e & \cite{Crandall1986}& 16\%              \\
         & e & \cite{Defrance1990}&7\%         \\
         & e & \cite{Donets1981}& 7\%             \\
\ion{Ne}{VIII}  & e & \cite{Riahi2001}&--\%           \\
\ion{Fe}{XXIV}  & t & \cite{Zhang1990}&--             \\
\ion{Zn}{XXVIII}  & t & \cite{Zhang1990}&--             \\[5pt]
 
\hline
 \caption{List of references for used cross-section data}
 
\end{tabular}
\begin{tablenotes}
    \item[1] e: experimental data, t: theoretical calculations
\end{tablenotes}
 \end{threeparttable}
 
\newpage
\begin{tabular}{l c  p{4cm} c}

\hline
\hline
Ion & Type\footnote[1] & Reference & Uncertainty\\[5pt]
\hline
& &  {\bfseries Be-sequence (2s)} \\ 
\hline
\ion{C}{III}    & e &  \cite{Fogle2008} & 8\%                     \\
\ion{N}{IV}     & e & \cite{Fogle2008} & 8\%                    \\
\ion{O}{V}      & e & \cite{Fogle2008} & 8\%                    \\
\ion{F}{VI}     & t & \cite{Younger1981d} &--   \\
\ion{Ne}{VII}   & t & \cite{Duponchelle1997} & 6\%  \\
\ion{S}{XIII}   & e & \cite{Hahn2012a} & 15\%  \\
\ion{Ar}{XV}    & t & \cite{Younger1981d} &--                    \\
\ion{Fe}{XXIII} & t &  \cite{Younger1982a} &--                    \\[5pt] 
\hline
& & {\bfseries B-sequence (2p)}  \\
\hline
\ion{B}{I}      & e & \cite{Tawara2002} (CHIANTI) & --                \\
\ion{C}{II}     & e & \cite{Aitken1971}&  7\%                \\
         & e & \cite{Yamada1989a}& 10\%               \\
\ion{N}{III}    & e & \cite{Aitken1971}& 7\%                \\
         & e & \cite{Bannister1996b}& 8\%    \\
\ion{O}{IV}     & e & \cite{Crandall1979}& 8\%         \\
\ion{Ne}{VI}    & e & \cite{Bannister1996a} & 11\%              \\   
         & e & \cite{Duponchelle1997} & 6\%              \\ 
\ion{Mg}{VIII}  & e & \cite{Hahn2010} & 15\%              \\    
\ion{Fe}{XXII}  & t & \cite{Zhang1990}& --              \\
\ion{Zn}{XXVI}  & t & \cite{Zhang1990} & --              \\[5pt]
 
\hline
& & {\bfseries C-sequence (2p)} \\
\hline
\ion{C}{I}      & e & \cite{Brook1978}& 5\%                 \\
\ion{N}{II}     & e & \cite{Yamada1989a}& 10\%                 \\
\ion{O}{III}    & e & \cite{Aitken1971}& 7\%             \\
         & e & \cite{Donets1981} & --          \\
         & e & \cite{Falk1980} & 10\%          \\
\ion{Ne}{V}    & e & \cite{Bannister1996a} & 9\%              \\    
         & e & \cite{Duponchelle1997} & 5\%              \\ 
\ion{Fe}{XXI}   & t & \cite{Zhang1990}& --           \\
\ion{Zn}{XXV}   & t & \cite{Zhang1990}& --           \\[5pt]
 
\hline
& & {\bfseries N-sequence (2p)} \\
\hline
\ion{N}{I}      & e & \cite{Brook1978}& 4\%                  \\
\ion{O}{II}     & e & \cite{Aitken1971}&  --            \\
                & e & \cite{Yamada1989a}& 7\%                 \\
\ion{F}{III}    & e & \cite{Muller1985a}&  9\%                \\
\ion{Ne}{IV}    & e & \cite{Gregory1983}&  8\%                \\
\ion{Si}{VIII}  & e & \cite{Zeijlmans1993}&7\%            \\
\ion{Fe}{XX}    & t & \cite{Zhang1990}& --           \\
\ion{Zn}{XXIV}  & t & \cite{Zhang1990}& --           \\[5pt]
 
\hline
& & {\bfseries O-sequence (2p)} \\
\hline
\ion{O}{I}      & e & \cite{Brook1978}& 5\%                    \\
                & e & \cite{Thompson1995}& 5\%                    \\
\ion{F}{II}     & e & \cite{Yamada1989b}& 10\%                 \\
\ion{Ne}{III}   & e & \cite{Bannister1996a}& 9\%   \\
\ion{Ar}{XI}    & e & \cite{Zhang2002}& 9\%                  \\
\ion{Fe}{XIX}   & t & \cite{Zhang1990}& --           \\
\ion{Zn}{XXIII} & t & \cite{Zhang1990}& --           \\[5pt]
 
\hline

\end{tabular}

\begin{tabular}{l c p{4cm} c}
\hline
\hline
Ion & Type\footnote[1] & Reference & Uncertainty \\[5pt]
\hline
& & {\bfseries F-sequence (2p)}\\
\hline
\ion{F}{I}      & e & \cite{Hayes1987}& 20\%                 \\
\ion{Ne}{II}    & e & \cite{Achenbach1984}& 10\%             \\
         & e & \cite{Diserens1984}& 3\%             \\
         & e & \cite{Donets1981} & --           \\
         & e & \cite{Man1987b}& 3\%                   \\
\ion{Al}{V}     & e & \cite{Aichele2001} & 8\%             \\
         & t & \cite{McGuire1982} & --            \\
\ion{Si}{VI}    & e & \cite{Thompson1994} & 4\%             \\
\ion{Fe}{XVIII} & e & \cite{Hahn2013}& 17\%           \\
         & t & \cite{Zhang1990}& --           \\
\ion{Zn}{XXII} & t & \cite{Zhang1990}& --           \\[5pt]
 
\hline
& & {\bfseries Ne-sequence (2s)}\\
\hline
\ion{Na}{II}    & e & \cite{Younger1981c}& --               \\
\ion{Mg}{III}   & e & \cite{Younger1981c}& --               \\
\ion{Al}{IV}    & e & \cite{Younger1981c}& --               \\
\ion{P}{VI}     & e & \cite{Younger1981c}& --               \\
\ion{Ar}{IX}    & e & \cite{Younger1981c}& --               \\
\ion{Fe}{XVII} & e & \cite{Zhang1990}& --               \\
\ion{Zn}{XXI}   & e & \cite{Zhang1990}& --  \\[5pt]
 
\hline
& & {\bfseries Ne-sequence (2p)}\\
\hline
\ion{Ne}{I}     & e & \cite{Almeida1995}& 8\%                  \\
         & e & \cite{Nagy1980}& 7\%                 \\ 
         & e & \cite{Stephan1980}& 8\%                 \\
         & e & \cite{Wetzel1987}& 15\%                  \\
\ion{Na}{II}    & e & \cite{Hirayama1986}& 13\%               \\
         & e & \cite{Hooper1966}& 10\%                \\
         & e & \cite{Peart1968}& 6\%             \\
\ion{Mg}{III}   & e & \cite{Peart1969}& 8\%                 \\
\ion{Al}{IV}    & t & \cite{Younger1981b}& --                 \\
\ion{Ar}{IX}    & e & \cite{Defrance1987}& 10\%            \\
         & e & \cite{Zhang2002}& 5\%            \\
\ion{Fe}{XVII}  & e & \cite{Hahn2013}& 16\%           \\
         & t & \cite{Zhang1990}& --            \\
\ion{Zn}{XXI}  & t & \cite{Zhang1990}& --   \\[5pt]
 
\hline
& & {\bfseries Na-sequence (2s \& 2p)} \\
\hline
\ion{Mg}{II}    & t & \cite{Younger1981c} &--               \\
\ion{Al}{III}   & t & \cite{Younger1981c} &--               \\
\ion{P}{V}      & t & \cite{Younger1981c} &--               \\
\ion{Ar}{VIII}  & t & \cite{Younger1981c} &--                  \\
\ion{Ni}{XVIII} & t & \cite{Pindzola1991} &--                  \\[5pt]

\hline
& & {\bfseries Na-sequence (3s)} \\
\hline
\ion{Na}{I}     & e & \cite{McFarland1965} & 8\%          \\  
         & e & \cite{Zapesochnyi1969}& 15\%      \\
\ion{Mg}{II}    & e & \cite{Becker2004}&10\%                \\
         & e & \cite{Martin1968}&11\%                \\
         & e & \cite{Peart1991b}&9\%                \\
\ion{Al}{III}   & e & \cite{Thomason1998}&8\%                  \\
\ion{Si}{IV}    & e & \cite{Crandall1982}&12\%                \\
\ion{Ti}{XII}   & e & \cite{Gregory1990}&7\%                   \\
         & t & \cite{Griffin1987}&--                   \\
\ion{Fe}{XVI}   & e & \cite{Gregory1987}&14\%                \\
         & e & \cite{Linkemann1995}&20\%                \\
\ion{Ni}{XVIII} & t & \cite{Pindzola1991}&--                \\[5pt]
 
\hline

\end{tabular}
\newpage
\begin{tabular}{l c p{4cm} c}
\hline
\hline
Ion & Type\footnote[1] & Reference & Uncertainty \\[5pt]
\hline
& & {\bfseries Mg-sequence (3s)} \\
\hline
\ion{Mg}{I}     & e & \cite{Boivin1998}&11\%               \\
         & e & \cite{Freund1990}&10\%                \\
         & e & \cite{McCallion1992a}&12\%                \\
\ion{Al}{II}    & e & \cite{Belic1987}&9\%               \\
\ion{Cl}{VI}    & e & \cite{Howald1986}&10\%               \\
\ion{Ar}{VII}   & e & \cite{Chung1996}&6\%                \\
         & e & \cite{Zhang2002}&7\%                \\
\ion{Sc}{X}     & t & \cite{Younger1983}&--                     \\
\ion{Fe}{XV}    & t & \cite{Bernhardt2014}&26\%                   \\[5pt] 
\hline
& & {\bfseries Al-sequence (3p)} \\
\hline
\ion{Al}{I}      & e & \cite{Freund1990}&10\%                \\
\ion{Si}{II}     & e & \cite{Djuric1993b}&9\%                \\
\ion{Cl}{V}      & e & \cite{Bannister1993}&9\%                \\
\ion{Ar}{VI}     & e & \cite{Gregory1982}&11\%                 \\
\ion{Sc}{IX}     & t & \cite{Younger1983}&--                 \\
\ion{Fe}{XIV}    & e & \cite{Hahn2013}&16\%                \\  [5pt]
 
\hline
& & {\bfseries Si-sequence (3p)}  \\
\hline
\ion{Si}{I}     & e & \cite{Freund1990}&10\%                   \\
\ion{P}{II}     & e & \cite{Yamada1989a}&10\%               \\
\ion{S}{III}    & e & \cite{Yamada1988}&10\%                 \\
\ion{Ar}{V}     & t & \cite{Arnaud1985}&--                       \\
\ion{Sc}{VIII}  & t & \cite{Younger1983}&--                       \\
\ion{Fe}{XIII}  & e & \cite{Hahn2012b}&12\%                     \\[5pt]
 
\hline
& & {\bfseries P-sequence (3p)}\\
\hline
\ion{P}{I}      & e & \cite{Freund1990}&10\%               \\
\ion{Cl}{III}   & e & \cite{Muller1985a}&10\%               \\
\ion{Sc}{VII}   & t & \cite{Younger1983}&--                     \\
\ion{Fe}{XII}   & e & \cite{Hahn2011a}&16\%                  \\
\ion{Ni}{XIV}   & e & \cite{Cherkani-Hassani2001}&14\%                  \\[5pt]
 
\hline
& &  {\bfseries S-sequence (3p)} \\
\hline
\ion{S}{I}      & e & \cite{Freund1990}&10\%                \\
\ion{Cl}{II}    & e & \cite{Djuric1993b}&7\%               \\
\ion{Ar}{III}   & e & \cite{Diserens1988}&3\%             \\
         & e & \cite{Man1993}&3\%             \\
\ion{Sc}{VI}    & t & \cite{Younger1983}&--                 \\
\ion{Fe}{XI}    & e & \cite{Hahn2012c}&9\%                        \\[5pt] 
\hline
& & {\bfseries Cl-sequence (3p)}\\
\hline
\ion{Cl}{I}     & e & \cite{Hayes1987}&14\%              \\
\ion{Ar}{II}    & e & \cite{Gao1997}&10\%             \\
         & e & \cite{Man1987b}&3\%                  \\
         & e & \cite{Muller1985a}&10\%            \\
         & e & \cite{Yamada1989b}& 10\%                   \\
\ion{K}{III}    & t & \cite{Younger1982c}&--                       \\
\ion{Sc}{V}     & t & \cite{Younger1982c}&--                      \\
\ion{Fe}{X}     & e & \cite{Hahn2012c}&9\%                        \\
\ion{Ni}{XII}   & e & \cite{Cherkani-Hassani2001}&14\%
\\[5pt]
 
\hline

\end{tabular}

\newpage
\begin{tabular}{l c p{4cm} c}
\hline
\hline
Ion & Type\footnote[1] & Reference & Uncertainty \\[5pt]
\hline
& &{\bfseries Ar-sequence (3p)}\\
\hline
\ion{Ar}{I}     & e & \cite{Ma1991}&15\%                   \\
         & e & \cite{McCallion1992b}&6\%            \\
         & e & \cite{Nagy1980}&6\%                  \\
         & e & \cite{Straub1995}&8\%                \\
         & e & \cite{Wetzel1987}&3\%                 \\
\ion{K}{II}     & e & \cite{Hirayama1986}&15\%            \\
         & t & \cite{Kumar1979}&--                  \\
         & e & \cite{Peart1968}&15\%                \\
\ion{Sc}{IV}    & t & \cite{Younger1982d}&--                        \\
\ion{Fe}{IX}    & e & \cite{Hahn2016}&16\%                        \\[5pt] 
\hline
& &{\bfseries K-sequence (3d \& 4s)}\\
\hline
\ion{K}{I}      & t & \cite{McCarthy1983}&--                  \\
\ion{Ca}{II}    & e & \cite{Peart1975}&10\% \\
         & e & \cite{Peart1989}&8\% \\ 
\ion{Sc}{III}  & e & \cite{Pindzola1994}&8\%\\ 
\ion{Ti}{IV}  & e & \cite{Falk1983c}&7\%\\   
\ion{Fe}{VIII}  & e & \cite{Hahn2015}&12\%\\     
\ion{Ni}{X}     & t & \cite{Pindzola1991}&--\\ [5pt]
 
\hline
& &{\bfseries Ca-sequence (3d \& 4s)}\\
\hline
\ion{Ca}{I}     & t & \cite{Roy1983}&--\\
         & t & \cite{McGuire1977}&-- \\
         & t & \cite{McGuire1997} &-- \\
\ion{Sc}{II}    & e & \cite{Jacobi2004} & 15\% \\
\ion{Ti}{III}   & e & \cite{Diserens1988}&3\% \\
         & e & \cite{Muller1985a}&9\% \\
\ion{Fe}{VII}   & e & \cite{Gregory1986}&5\%\\  
         & e & \cite{Stenke1999}&8\%\\
\ion{Ni}{IX} & t & \cite{Pindzola1991}&--\\ 
             & e & \cite{Wang1988}&6\%\\[5pt]
 
\hline
& &{\bfseries Sc-sequence (3d \& 4s)}\\
\hline
\ion{Sc}{I}    & e & \cite{Tawara2002} (CHIANTI)& -- \\
\ion{Ti}{II}   & e & \cite{Diserens1988}&3\% \\
\ion{Fe}{VI}   & e & \cite{Gregory1986}&5\%\\  
        & e & \cite{Stenke1999}&8\%\\
\ion{Ni}{VIII} & t & \cite{Pindzola1991}&--\\ 
        & e & \cite{Wang1988}&6\%\\  [5pt]
 
\hline
& &{\bfseries Ti-sequence (3d \& 4s)}\\
\hline
\ion{Ti}{I}    & t & \cite{McGuire1977}&-- \\
\ion{Fe}{V}    & e & \cite{Stenke1999}&8\%\\
\ion{Ni}{VII}  & e & \cite{Wang1988}&6\%\\ [5pt] 
 
\hline
& &{\bfseries V-sequence (3d \& 4s)}\\
\hline
\ion{V}{I} & e & \cite{Tawara2002} (CHIANTI) & --  \\
\ion{Cr}{II}   & e & \cite{Man1987a}&2.5\% \\
\ion{Fe}{IV}   & e & \cite{Stenke1999}&--\\
\ion{Ni}{VI}   & e & \cite{Wang1988}&6\%\\  [5pt]
 
\hline
& &{\bfseries Cr-sequence (3d \& 4s)}\\
\hline
\ion{Cr}{I}     & t & \cite{Reid1992}&-- \\
                  & t & \cite{McGuire1977}&-- \\
\ion{Fe}{III}   & t & FAC (CHIANTI) &--\\  
\ion{Ni}{V}    & e & \cite{Bannister1993}&7\%\\ 
        & t & \cite{Pindzola1991}&--\\ [5pt]
 
\hline

\end{tabular}

\newpage
\begin{tabular}{l c p{3.5cm} c}
\hline
\hline
Ion & Type\footnote[1] & Reference & Uncertainty \\[5pt]
\hline
& &{\bfseries Mn-sequence (3d \& 4s)}\\
\hline
\ion{Mn}{I}  & e & \cite{Tawara2002} (CHIANTI) & -- \\
\ion{Fe}{II}   & t & \cite{Younger1983} (3d) &--\\
              & e & \cite{Montague1984b} (4s)&-- \\ 
\hline
& &{\bfseries Fe-sequence (3d \& 4s)}\\
\hline
\ion{Fe}{I}  & e & \cite{Freund1990}&7\%\\ 
\ion{Co}{II}  & t & FAC (CHIANTI) &--\\ 
\ion{Ni}{III}  & t & \cite{Pindzola1991}&--\\ 
\ion{Cu}{IV}   & e & \cite{Gregory1986}&4\%\\ [5pt]
\hline
& &{\bfseries Co-sequence (3d \& 4s)}\\
\hline
\ion{Co}{I} & e & \cite{Tawara2002} (CHIANTI) & --                \\
\ion{Ni}{II}  & e & \cite{Montague1984a}&3\%\\ 
\ion{Cu}{III}   & e & \cite{Gregory1986}&4\%\\[5pt] 
 
\hline
& &{\bfseries Ni-sequence (3d \& 4s)}\\
\hline
\ion{Ni}{I}   & t & \cite{Pindzola1991} (3d)&-- \\
       & t & \cite{McGuire1977} (4s)&-- \\ [5pt]

\hline
& &{\bfseries Cu-sequence (3d \& 4s)}\\
\hline
\ion{Cu}{I}  & t & FAC (3d) &-- \\
      & e & \cite{Bolorizadeh1994} (4s)&10\%\\
      & t & \cite{Bartlett2002} (4s)&--\\
\ion{Zn}{II} & t & FAC (3d) &-- \\
      & e & \cite{Peart1991a} (4s)&10\%\\ 
      & e & \cite{Rogers1982} (4s)&10\%\\ [5pt]
\hline
& &{\bfseries Zn-sequence (3d \& 4s)}\\ 
\hline
\ion{Zn}{I}  & t & FAC (3d) &-- \\
      & t & \cite{McGuire1977} (4s) &-- \\
      & t & \cite{Omidvar1977} (4s)&--\\ [5pt]
\hline

\end{tabular}

\newpage

\section{Calculation of DI ionization rate coefficients}

As explained in Section 2, for the direct ionization cross-section calculation, the extended Younger's equation (\ref{eqn:younger_ext}) was  used:\begin{equation}
\displaystyle uI^2Q_{DI} = A\left(1-{1\over u}\right) + B\left(1 - {1\over u}\right)^2 + C R \ln
u + D {\ln u\over \sqrt{u}} +  E{\ln u\over u},
\label{eqn:younger}
\end{equation}

where
$$R\simeq 1+1.5\epsilon+0.25\epsilon^2$$ $$\epsilon \equiv E/mc^2 \equiv uI/mc^2$$ $$I/mc^2 \equiv \lambda\ll1$$ $$u=E/I$$

with E the kinetic energy of the colliding electron and I the ionisation potential of the relevant subshell.

If we generalize the formula for all the inner shells $j$ and the summation over all shells is taken into account for the total direct ionization cross-section, the parametric formula is
\begin{multline}
u_jI_j^2Q_{DI} =\sum_j\bigg[A_j\left(1-{1\over u_j}\right) + B_j \left(1 - {1\over u_j}\right)^2 \\+ C_j R_j \ln u_j  + D_j {\ln u_j\over \sqrt{u_j}} +  E_j{\ln u_j\over u_j}\bigg].
\end{multline}

Eq. \ref{eqn:younger} can be written as follows being $u_j=E/I_j$:
\begin{align}
\centering
u_jI_j^2Q_{DI} =\sum_{i=1}^8 c_i \cdot f_i(u_j),
\end{align}

with

\begin{tabular}{l l}
\\
$c_1=(A_j + B_j)$  &             $f_1(u)=1$\\
$c_2=(-A_j - 2B_j)$&          $   f_2(u)={1\over u}$\\
$c_3=B_j $ &                    $ f_3(u)={1\over u^2}$\\
$c_4=C_j $  & $                   f_4(u)=\ln u$\\
$c_5={3\over{2}}\lambda C_j$ &  $ f_5(u)=u\ln u$\\
$c_6={1\over{4}}\lambda^2 C_j$& $ f_6(u)=u^2\ln u$\\
$c_7=D_j  $  &                  $ f_7(u)={{\ln u}\over{\sqrt{u}}}$\\
$c_8=E_j  $   &  $                f_8(u)={{\ln u}\over{u}}$\\
\\
\end{tabular}

As a consequence, the direct ionization rate coefficients versus the temperature [T] are:
\begin{equation}
\displaystyle C_{DI} =r_0\int_1^\infty(u_jI_j^2Q_{DI})e^{-u_jy}\mathrm{d}u_j \equiv r_0\sum_{i\thickapprox 1}^\infty c_i \cdot g_i(u_j) ,
\label{eqn:ir_di_1}
\end{equation}

with $y\equiv{I/kT}$
and $ r_0 \equiv {{2\sqrt{2}n_en_i}\over{[\pi(kT)^3m_e]^{1\over2}}}$
\begin{equation}
\displaystyle  g_i(y) = \int_1^\infty f_i(u) e^{-uy} \mathrm{d}u\\
\end{equation}

with

\begin{align}
g_1(y)&=\int_1^\infty e^{-uy} \mathrm{d}u={1\over y}e^{-y}\\
g_2(y)&=\int_1^\infty \frac {e^{-uy}}{u} \mathrm{d}u=E_1(y)\\
\end{align}
being $E_1$ the first exponential integral function.
\begin{align}
g_3(y)&=\int_1^\infty \frac {e^{-uy}}{u^2} \mathrm{d}u=e^{-y}-yE_1(y)\\
g_4(y)&=\int_1^\infty \ln{u} e^{-uy} \mathrm{d}u={1\over y}E_1(y)\\
g_5(y)&=\int_1^\infty u\ln{u} e^{-uy} \mathrm{d}u={1\over y^2}[e^{-y}+E_1(y)]\\
g_6(y)&=\int_1^\infty u^2\ln{u} e^{-uy} \mathrm{d}u={{3+y}\over y^3}e^{-y}+{2\over y^3}E_1(y)\\
g_7(y)&=\int_1^\infty \frac{\ln{u}}{\sqrt{u}} e^{-uy} \mathrm{d}u
\end{align}

For small y (y<0.6):
\begin{equation}
\displaystyle g_7(y)\simeq-\sqrt{\pi\over y}(\gamma+\ln 4+ \ln y)+4-\frac{4y}{9}+ \frac{2y^2}{25}-\frac{2y^3}{147}+\frac{y^4}{486},
\end{equation}

where $\gamma$=0.577216 is the Euler-Mascheroni constant.\\

For intermediate y $({0.6\leq y \geq {20}})$:
\begin{equation}
\displaystyle g_7(y)\simeq\frac{e^{-y}[p_1+\frac{p_2}{\sqrt{y}}+ \frac{p_3}{y}+\frac{p_4\ln y}{\sqrt{y}}]}{(y+p_5)(y+p_6)} 
\end{equation}
with,
\begin{align*}
p_1 &=1.000224  \\
p_2 &=-0.11301  \\
p_3 &=1.851039 \\
p_4 &=0.019731\\
p_5 &=0.921832\\
p_6 &=2.651957
\end{align*}

For large y (y>20):
\begin{equation}
\displaystyle g_7(y)\simeq \frac{e^{-y}}{y^2}[1-\frac{2}{y}+ \frac{23}{4y^2}-\frac{22}{y^3}+\frac{1689}{16y^4}-\frac{4881}{8y^5} ]
\end{equation}

\begin{align}
g_8(y)&=\int_1^\infty \frac{\ln{u}}{u} e^{-uy} \mathrm{d}u
\end{align}

For small y (y<0.5):
\begin{equation}
\displaystyle g_8(y)\simeq\gamma\ln y+ {1\over2}(\ln y)^2-y+\frac{y^2}{8}- \frac{y^3}{54}+\frac{y^4}{384}-\frac{y^5}{3000}+0.989056
\end{equation}

For intermediate y $({0.5\leq y \geq {20}})$:
\begin{equation}
\displaystyle g_8(y)\simeq \frac{e^{-y}[a_o+\frac{a_1}{{y}}+ \frac{a_2}{y^2}+\frac{a_3}{{y^3}}+\frac{a_4}{y^4}]}{(y+b_1)(y+b_2)} 
\end{equation}
with,
\begin{align*}
a_0 &=0.999610841 \\
a_1 &=3.50020361 \\
a_2 &=-0.247885719 \\
a_3 &=0.00100539168\\
a_4 &=1.3907539.10^{-3}\\
b_1 &=1.84193516\\
b_2 &=4.64044905
\end{align*}

For large y (y>20):
\begin{equation}
\displaystyle g_8(y)\simeq \frac{e^{-y}}{y}\sum_{n=1}^7{\frac{a_n}{y^n}}
\end{equation}
with,
\begin{align*}
a_1 &=1 \\
a_2&=-3 \\
a_3 &=11 \\
a_4 &=-50 \\
a_5 &=274\\
a_6 &=-1764\\
a_7 &=13068
\end{align*}

\section{Calculation of EA ionization rate coefficients}

The excitation-autoionization ion rate coefficients were  calculated applying the integral to a Maxwellian velocity distribution of Mewe's equation, mentioned in  Section 2.2: 
\begin{equation}
uI_{EA}^2Q_{EA} =\bigg[A_{EA}+ {B_{EA}\over u}  + {C_{EA} \over u^2} + {2D_{EA} \over u^3} + E_{EA} \ln u\bigg]. \nonumber
\end{equation}

The EA cross-section contribution that affects the outer shell of each element, is the summation over $k$ energy level transitions with $I_{EAk}$ the excitation-autoionization potential being $u_k=E/I_EAk$:

\begin{equation}
u_kI_{EAk}^2Q_{EA}=\sum_{i=1}^5 d_i \cdot l_i(u_k),
\end{equation}

with,

\begin{tabular}{l l}
\\
$d_1=A_{EAk}$  &     $        l_1(u_k)=1$\\
$d_2=B_{EAk}$&            $ l_2(u_k)={1\over u_k}$\\
$d_3=C_{EAk} $ & $                    l_3(u_k)={1\over u^2_k}$\\
$d_4=2D_{EAk}  $&                  $  l_4(u_k)={1\over u^3_k}$\\
$d_5=E_{EAk} $ & $                   l_5(u_k)=\ln u_k$\\
\\
\end{tabular}

Therefore, the EA ionization rate coefficients versus the temperature [T] are
\begin{equation}
\displaystyle C_{EA} =r_0\int_1^\infty(u_kI_{EAk}^2Q_{EA})e^{-u_ky} \mathrm{d}u \equiv r_0\sum_{i=1}^5 d_i \cdot m_i(u_k),
\label{eqn:ir_EA_1}
\end{equation}

with $y\equiv{I_{EA}/kT}$
and $ r_0 \equiv {{2\sqrt{2}n_en_i}\over{[\pi(kT)^3m_])^{1\over2}}}$
\begin{align}
\displaystyle
 m_i(y) &= \int_1^\infty l_i(u_k) e^{-u_ky} \mathrm{d}u,
\end{align}

with:
\begin{align}
m_1(y)&=\int_1^\infty e^{-uy} \mathrm{d}u ={1\over y}e^{-y}\\
m_2(y)&=\int_1^\infty \frac {e^{-uy}}{u} \mathrm{d}u=E_1(y)\\
m_3(y)&=\int_1^\infty \frac {e^{-uy}}{u^2} \mathrm{d}u=e^{-y}-yE_1(y)\\
m_4(y)&=\int_1^\infty \frac {e^{-uy}}{u^3} \mathrm{d}u=(1-y){e^{-y}\over2}+{y^2\over2}E_1(y)\\
m_5(y)&=\int_1^\infty \ln{u} e^{-uy} \mathrm{d}u={1\over y}E_1(y)
\end{align}

\section{The DI coefficients}

The table below shows an example of the DI coefficients calculated by Eq. 
\ref{eqn:younger_ext} for Si-like \ion{Fe}{XI}.

\small
 \begin{threeparttable}
\begin{tabular}{c c c c c c c c c}
\hline
\hline
ii\tnote{a} & ir\tnote{b} & iz\tnote{c} & I$_{DI}$ (eV)\tnote{d}&A&B&C&D&E\tnote{e}\\[5pt]
\hline
16&1&26& 7585.000&28.28&-11.62&4.80&0.00&-24.12\\
16&2&26& 1164.000&18.21&-3.73&3.56&-3.85&-9.85\\
16&3&26& 1048.687&59.57&-26.85&13.230&14.61&-51.29\\
16&4&26& 324.400&21.91&-11.03&2.25&4.09&-18.89\\
16&5&26& 290.300&80.28&-72.24&6.22&33.59&-92.87\\[5pt]
\hline
 \caption{The DI coefficients}
\end{tabular}

\begin{tablenotes}
    \item[a]ii: Isoelectronic sequence 
    \item[b]ir: Shell number 1-7 (1s,2s,2p,3s,3p,3d,4s)
    \item[c]iz: Element
     \item[d] I$_{DI}$: Ionization potential
    \item[e]A,B,C,D and E units: $10^{-24}$~m$^2$keV$^2$
\end{tablenotes}
\end{threeparttable}

\section{The EA coefficients}

The table below shows an example of the DI coefficients calculated by equation 
(\ref{eqn:eamewe}) for Si-like \ion{Fe}{XI}.

\small
 \begin{threeparttable}
\begin{tabular}{c c c c c c c c c}
\hline
\hline

ii\tnote{a} & iz\tnote{b} & k\tnote{c} & I$_{EA}$ (eV)\tnote{d}&A&B&C&D&E\tnote{e}\\[5pt]
\hline
16 &26&  1 &  757.000 &     -0.465   &    0.812 &     -0.037   &   -0.062 &     0.608\\
16 &26&  2 &  802.500   &    2.809  &    -4.408  &     4.904  &    -1.017  &    -0.056\\
16 &26&  3  & 902.200   &    0.260   &   -0.062  &     0.006 &      0.006  &     0.000\\
16 &26&  4 &  916.900 &     -0.136      & 0.223       &0.248  &    -0.087     &  0.378\\
16 &26 & 5 &  936.300   &    0.372   &   -0.409    &   0.434  &    -0.068     &  0.000\\
16 &26 & 6 &  977.500  &     0.217   &   -0.143    &   0.322   &   -0.062      & 0.143\\
16 &26 & 7  & 662.900  &     1.302   &   -0.484    &   0.794   &   -0.174   &    0.006\\
16 &26 & 8 &  709.000  &    -0.831 &      7.738  &     5.871  &    -2.344 &      9.653\\
16 &26 & 9 &  809.300  &     0.186   &   -0.068   &    0.409 &     -0.099   &    0.093\\
16 &26 &10 &  823.800  &     0.955   &   -0.384  &     0.818   &   -0.149     &  0.000\\
16 &26& 11  & 843.400   &    1.048   &   -0.651     &  2.883 &     -0.670 &      1.445\\
16 &26 &12 &  884.700 &      0.942    &  -0.725  &    1.910    &  -0.378   &    0.521\\[5pt]

\hline

 \caption{The EA coefficients}
\end{tabular}

\begin{tablenotes}
    \item[a]ii: Isoelectronic sequence 
    \item[b]iz: Element
    \item[c]k: Number of transitions
     \item[d] I$_{EA}$: Ionization potential
    \item[e]A,B,C,D and E units: $10^{-24}$~m$^2$keV$^2$
\end{tablenotes}
\end{threeparttable}

\end{appendix}

\end{document}